\documentstyle[aaspp4,psfig]{article}

\newcommand{\FP}{\mbox {Fokker-Planck}}
\newcommand{\Aa}{Aa}
\newcommand{\Ae}{Ae}
\newcommand{\family}{\mbox{family}}
\newcommand{\nbody}{\mbox {$N$-body}}

\newcommand{\seba}{\mbox {${\sf SeBa}$}}
\newcommand{\msun}{\mbox {$M_\odot$}}
\newcommand{\aesc}{\mbox {$\nu_{\rm esc}$}}
\newcommand{\rt}{\mbox {$r_{\rm t}$}}
\newcommand{\sisu}[1]{$\times 10^{#1}$}

\def\apgt{\ {\raise-.5ex\hbox{$\buildrel>\over\sim$}}\ }
\def\aplt{\ {\raise-.5ex\hbox{$\buildrel<\over\sim$}}\ }

\slugcomment{Accepted for publication in ApJ}

\lefthead{K.\ Takahashi \& S.\ F.\ Portegies Zwart}
\righthead{The evolution of globular clusters in the Galaxy}

\begin{document}

\title{
    The Evolution of Globular Clusters in the Galaxy
}

\author{Koji Takahashi 
\footnote{Present address: Department of Astronomy, School of Science,
The University of Tokyo, 7-3-1 Hongo, Bunkyo-ku, Tokyo 113-0033, Japan}}
\affil{Department of Earth Science and Astronomy,\\
       College of Arts and Sciences, The University of Tokyo,\\
       3-8-1 Komaba, Meguro-ku, Tokyo 153-8902, Japan\\
    takahasi@chianti.c.u-tokyo.ac.jp}

\and

\author{Simon F. Portegies Zwart
\footnote{Hubble Fellow}}
\affil{
Department of Astronomy,
Boston University,\\
725 Commonwealth Avenue, Boston, MA 01581\\
spz@komodo.bu.edu}

\date{}


\begin{abstract}

We investigate the evolution of globular clusters using \nbody\
calculations and anisotropic \FP\ calculations.  The models include a
mass spectrum, mass loss due to stellar evolution, and the tidal field
of the parent galaxy.  Recent \nbody\ calculations have revealed a
serious discrepancy between the results of \nbody\ calculations and
isotropic \FP\ calculations.  The main reason for the discrepancy is
an oversimplified treatment of the tidal field employed in the {\em
isotropic} Fokker-Planck models.  In this paper we perform a series of
calculations with {\em anisotropic} \FP\ models with a better
treatment of the tidal boundary and compare these with \nbody\
calculations.  The new tidal boundary condition in our \FP\ model
includes one free parameter.  We find that a single value of this
parameter gives satisfactory agreement between the \nbody\ and \FP\
models over a wide range of initial conditions.

Using the improved \FP\ model, we carry out an extensive survey of the
evolution of globular clusters over a wide range of initial
conditions varying the slope of the mass function, the central concentration,
and the relaxation time.  The evolution of clusters is followed up to
the moment of core collapse or the disruption of the clusters in the
tidal field of the parent galaxy.  In general, our model clusters,
calculated with the anisotropic \FP\ model with the improved treatment
for the tidal boundary, live longer than isotropic
models.  The difference in the lifetime between the isotropic and
anisotropic models is particularly large when
the effect of mass loss via stellar evolution
is rather significant.
On the other hand the difference
is small for relaxation-dominated clusters which initially have steep
mass functions and high central concentrations.

\end{abstract}


\section{Introduction}

The most reliable method to model the evolution of dense star clusters
is the direct integration of the equations of motions of all stars
via \nbody\ calculations.  The high cost of such 
calculations, however, limits the choice of the number of stars to a
few ten thousands, a small number compared to real globular clusters,
and also limits the number of calculations which can be performed on
a practical time scale.

In the last few decades Fokker-Planck models have gained in popularity
for studying the dynamics of globular clusters (see Meylan \& Heggie
1997 for a recent review).  The relatively low computational cost of
Fokker-Planck models makes it possible to perform a large number of
calculations over broad ranges of initial parameters. This open the
possibility to study the initial parameter space which may have been
available at the formation of the population of observed globular
clusters.

On the other hand,
one has to pay for the higher speed of \FP\ calculations with a less
accuracy of these models. 
It is therefore important to examine
the accuracy of \FP\ models.
One way of such examinations is to do a comparative study between
\FP\, and \nbody\, calculations.

\subsection{What happened before}

The development of approximate methods for modeling the dynamical
evolution of star clusters started in the 1970's using Monte-Carlo
methods (Spitzer 1987, and references therein). Several groups applied
these methods to address various problems in stellar dynamics.  Cohn
(1979, 1980) developed a new method in which the orbit-averaged \FP\
equation was solved directly using a finite-difference technique.
Cohn (1980) demonstrated that his code could follow the evolution of
the cluster into much more advanced stages of core collapse than 
Monte-Carlo codes (see, however, Giersz 1998, for a possible revival
of Monte-Carlo methods).

In Cohn's calculations globular clusters were modeled as fairly idealized
systems, i.e.: isolated systems of identical point masses (Cohn 1979,
1980).  This is quite a natural starting point for theoretical
studies.  After Cohn's seminal works, many contributions have been
made to improve Cohn's \FP\ scheme in terms of adding various
astrophysical processes, such as a mass spectrum, primordial binaries,
a tidal cut-off, gravitational shocks by the galactic disk and bulge,
mass loss from stellar evolution, anisotropy of the velocity
distribution, rotation of the clusters, etc. (see Meylan \& Heggie
1997).

An important landmark in modeling the evolution of globular clusters
with \FP\ calculations was set by Chernoff \& Weinberg (1990,
hereafter CW).  Their models included a tidal cut-off by the
parent galaxy, a stellar mass spectrum, and mass loss due to stellar
evolution.

CW boldly added simultaneously several new astrophysical phenomena to
their models.  However, generally speaking,
each time a model is extended trying to make
simulations more ``realistic'', the number of assumptions usually
increases; therefore thorough testing is required (not only for \FP\ models) to
eliminate the possible introduction of omissions and the violation of
previously-made assumptions.  It is not easy to assess the reliability of a
model. The most reliable way to do this may well be detailed
comparison with another model which is based on a completely different
numerical method. 

Such comparisons have found good agreement between \nbody, \FP, and
gaseous models for isolated star clusters of point masses (Giersz
and Heggie 1994a, 1994b; Giersz and Spurzem 1994; Spurzem and
Takahashi 1995).  These are important results because they demonstrate
that two-body relaxation is a main physical process
which drives the dynamical evolution of star
clusters, as was expected from theory (see, e.g., Spitzer 1987), 
and therefore that the
time evolution of such relatively simple systems scales with the
two-body relaxation time.

Fukushige \& Heggie (1995) and Portegies Zwart et al.\ (1998)
demonstrated that \nbody\ calculations give significantly different
results from the isotropic \FP\ calculations of CW for some initial conditions.
These \nbody\ and \FP\ calculations included the steady tidal field of the
parent galaxy and mass loss due to the evolution of single stars.
The results of Portegies Zwart et al.\ (1998) can be summarized as
follows: 1) The clusters simulated with the $N$-body model live more
than an order of magnitude longer than the comparable isotropic \FP\
model; 2) The lifetime of the clusters depends on the number of stars
$N$ in a rather complex way. The clusters with a larger number of stars
live shorter contrary to what was expected).  A similar $N$-dependence
of \nbody\ models was also found in a collaborative experiment
organized by Heggie (Heggie et al.\ 1999).

Takahashi \& Portegies Zwart (1998, TPZ hereafter) solved the
discrepancies pointed out by Portegies Zwart et al.\ (1998) by using
more general {\it anisotropic} \FP\ models
together with an improved treatment of the tidal boundary.
The improved \FP\ models are in
in excellent agreement with \nbody\ models.

The assumption of velocity isotropy and the oversimplified escape
condition adopted by CW turned out to result in an enormous
overestimate of the escape rate, i.e.: an underestimate of the life
time of the cluster.
TPZ took account of
the orbital angular momentum as well as the orbital energy
to determine escaping stars.
TPZ also took account of that 
a star on an escape orbit requires some time to
leave the cluster.  The number of stars that escape per relaxation
time depends therefore on the ratio of the relaxation time to the
dynamical time, i.e.: on the number of stars.

TPZ's implementation of the escape condition adds a free parameter
\aesc\ (denoted by $\alpha_{\rm esc}$ in TPZ) to \FP\,
models. This parameter determines the time scale on which escaping
stars leave the cluster, and should be of the order of the crossing
time at the tidal radius (Lee \& Ostriker 1987; see also Ross et al.\ 
1997).  TPZ compared \FP\, calculations with \nbody\, calculations to
calibrated \aesc\ for one specific set of initial conditions: density
profile, mass function and relaxation time.  However, it is not
obvious that a single value of \aesc\ gives good agreement between
Fokker-Planck and \nbody\ models for a wide range of initial
conditions. To test this hypothesis is the first main purpose of this
paper.

The second purpose of this paper is to carry out an extensive survey
of the evolution of globular clusters in the
Galaxy for various initial conditions. 
The same initial conditions will be used as those of the
survey by CW.  We perform the survey with the
improved anisotropic \FP\ models with the calibrated value for \aesc.

In the following section we review the anisotropic Fokker-Planck model
and the \nbody\ model. In section \ref{sec:stellar} the stellar
evolution models are described, and the initial conditions are reviewed
in section \ref{sec:initial}.  The results of a comparison
between \FP\ and \nbody\ models are presented in section \ref{sec:FPvsNB}.
\FP\ survey results are shown in section \ref{sec:FPsurvey}.
Section \ref{sec:discussion} discusses
our simulation results, including
comparison with other \nbody\ simulations and with observations. 
We summarize our findings in section\,\ref{conclusions}.

\section{The Models}\label{sec:models}

\subsection{The \FP\ model}\label{sec:fp}

We use anisotropic orbit-averaged Fokker-Planck models (Cohn 1979;
Takahashi 1995).  The star cluster is assumed to be spherically
symmetric and in dynamical equilibrium at any time.  The distribution
function $f$ at time $t$ is then a function of the energy of a star
per unit mass, $E$, and the angular momentum per unit mass, $J$ (see
e.g., Spitzer 1987, section 1.2).  On the other hand, isotropic \FP\,
models (Cohn 1980) assume that 
the distribution function does not depend
on the angular momentum, i.e., that the velocity distribution is
isotropic at every point in the cluster.

Anisotropic \FP\ models are more general than isotropic \FP\ models
and therefore preferable.  The higher computational costs and
more complex numerical implementation, however, caused anisotropic
\FP\ models to be less favored than their isotropic cousins (Cohn
1980).  The numerical problem which was encountered by Cohn (1979)
in his anisotropic \FP\ calculations was solved by Takahashi (1995;
1996) mainly by implementing a new finite-difference scheme (see also Drukier
et al.\ 1999).

All \FP\ calculations in this paper are performed with the code which
was developed by Takahashi (1995; 1996) and was later modified to include
a mass function for single stars (Takahashi 1997), a tidal boundary
(Takahashi et al.\ 1997; TPZ), and stellar evolution (TPZ).

\subsubsection{Determination of escaping stars}\label{Sect:escape}

TPZ demonstrated that the results of \FP\ calculations are sensitive
to the choice of the tidal boundary conditions.  Two different
criteria for the identification of escapers were tested by Takahashi
et al.\ (1997) and by TPZ. These are: 1) the apocenter criterion, in
which a star is removed if
\begin{equation}
r_{\rm a}(E,J) > r_{\rm t} ;
\end{equation}
and 2) the energy criterion, in which a star is removed if
\begin{equation}
E > E_{\rm t} \equiv -GM/r_{\rm t}.
\end{equation}
Here $r_{\rm t}$ is the tidal radius of the cluster,
$r_{\rm a}(E,J)$ is the apocenter distance of a star with energy
$E$ and angular momentum $J$, $G$ is the gravitational constant,
and $M$ is the cluster mass.

The apocenter criterion is considered more realistic than the energy
criterion.  The latter tends to remove too many stars too quickly,
since it
may even remove stars orbiting inside the tidal radius.  For most
calculations we adopt the {\bf a}pocenter criterion to determine
escapers for {\bf A}nisotropic \FP\ models; we call these models \Aa\ models.
We also adopt the {\bf e}nergy criterion for some calculations, these
models are called \Ae\ models.

\subsubsection{Introducing the crossing time in the \FP\ model}

In the \FP\ calculations performed by CW, escapers are removed
instantaneously from the clusters.  In the absence of stellar
evolution this condition may be justified if the dynamical time scale
of the cluster is negligible compared to the two-body relaxation time,
i.e.; if the number of stars $N \rightarrow \infty$.

TPZ demonstrated that the lifetime of stellar
systems with small $N$ is dramatically affected by the time scale on
which escaping stars are removed from the cluster.  Two treatments of
escaping stars were tested by TPZ: 1) instantaneous escape: a star is
removed from the cluster as soon as it satisfies an adopted escape
criterion; 2) crossing-time escape: a star on an escape orbit
hangs around in the cluster
for a certain time before it really leaves the cluster.

To implement the crossing-time escape condition in \FP\
calculations we use the boundary condition of Lee \& Ostriker
(1987):
\begin{equation}
     \frac{\partial f}{\partial t} = - \aesc f \left[ 1-
                                   \left\vert
	                                \frac{E}{E_{\rm t}}
	                           \right\vert^3
                               \right]^{1/2} 
                       \frac{1}{2\pi}\sqrt{
                       \frac{4\pi}{3}G\rho_{\rm t}}.
\label{eq:LO}\end{equation}
Here $\rho_{\rm t}$ is the mean mass density within the tidal radius
and \aesc\ is a dimensionless constant which determines the time scale
on which escapers leave the cluster.  This equation is applied in a
region of phase space where the adopted escape criterion is satisfied.
Eq.~(\ref{eq:LO}) is derived assuming that
escapers leave the cluster on a dynamical time scale.  For $N
\rightarrow \infty$ with a fixed relaxation time, this condition
becomes identical to removing escapers instantaneously, since the
ratio of the crossing time to the relaxation time vanishes.

When Eq.~(\ref{eq:LO}) is used, the density at the
tidal radius is generally still finite.  The total mass of the
cluster is then defined by the mass within $r_{\rm t}$.

\subsubsection{The mass function} \label{FPmassfunction}

In our \FP\ models a continuous mass function is represented by $K$
discrete mass components.  We use $K=20$ components for all \FP\
calculations.  This number is considered to be sufficiently large for
our calculations (see CW).  At $t=0$ the mass of a star in mass-bin
$k$ is given by
\begin{equation}
	m_k = m_-\left(\frac{m_+}{m_-}
                       \right)^{(k-\frac{1}{2})/K}.
\end{equation}
Here $m_-$ and $m_+$ are the lower and upper limits of the initial
mass function.  The total mass in mass interval $k$ is
\begin{equation}
M_k=\int_{m_{k-\frac{1}{2}}}^{m_{k+\frac{1}{2}}} \frac{dN}{dm} m \,dm \,.
\end{equation}
Here $dN/dm$ is the number of stars per unit mass interval.

\subsection{The \nbody\ model}

The \nbody\ portion of the simulations is carried out using the {\tt
Kira} integrator, operating within the Starlab (version 3.3) software
environment (McMillan \& Hut 1996; Portegies Zwart et al.\
1998).\nocite{mh96} Time integration of stellar orbits is accomplished
using a fourth-order Hermite scheme (Makino \& Aarseth
1992).\nocite{ma92} {\tt Kira} also incorporates block time steps
(McMillan 1986a; 1986b; Makino
1991),\nocite{1986ApJ...307..126M}\nocite{1986ApJ...306..552M}
\nocite{1991ApJ...369..200M} special treatment of close two-body and
multiple encounters of arbitrary complexity, and a robust treatment of
stellar and binary evolution and stellar collisions (Portegies Zwart
et al.\ 1999).  The special-purpose GRAPE-4 (Makino et al.\
1997)\nocite{1997ApJ...480..432M} system is used to accelerate the
computation of gravitational forces between stars.  The treatment of
stellar mass loss is described in Portegies Zwart et al.\ 
(1998).\nocite{1998A&A...337..363P} A more complete description of the
Starlab environment is in preparation, but more information can be
found at the following URL: {\tt http://www.ias.edu/\~\/starlab}.

\section{Stellar Evolution}\label{sec:stellar}

\subsection{\nbody\ Models}

Stars in the $N$-body calculations are evolved with the stellar
evolution model \seba\footnote{The name {\sf SeBa} is adopted from the
Egyptian word for `to teach', `the door to knowledge' or `(multiple)
star'. The exact meaning depends on the hieroglyphic spelling.}
(Portegies Zwart \& Verbunt 1996, Section 2.1).\nocite{pzv96} In
\seba\ stars with mass $ m > 8\msun$ become neutron stars
with mass $m_{\rm fin}=1.34$\,\msun, and lower mass stars become white
dwarfs.  The mass of these white dwarfs are given by the core mass of
their progenitors.

The evolution of the stars in the \FP\ calculations which are compared
with the \nbody\ runs is also computed using \seba.
In practice tabulated data as listed in Table \ref{tab:se} are used.

\subsection{\FP\ models}

We use the same stellar evolution model as was adopted by CW when we
compare our anisotropic Fokker-Planck models with CW's models.
The stars in this stellar evolution model live somewhat shorter than those in
\seba\ (see Table \ref{tab:se}).

\placetable{tab:se}

Table \ref{tab:se} compares the stellar evolution model used by CW
with \seba.  The main-sequence lifetimes in the CW model are obtained
by fitting cubic splines to the data listed in Table 1 of CW.  They do
not mention the lifetimes of stars with mass $m \leq 0.83 \msun$.
Some clusters become considerably older than 15\,Gyr and it is
necessary to use proper lifetimes for these stars as well.
We assume that the lifetime of stars with $m \le 0.83 \msun$ is
proportional to $m^{-3.5}$ (Drukier 1995).  This extrapolation gives
also a good fit to the lifetime of the low-mass stars in \seba\ (see
Table \ref{tab:se}).

At the end of the main-sequence life, a star with an initial mass
$m_{\rm ini}$ forms a compact object with mass $m_{\rm fin}$.  In the
model adopted by CW the remnant mass is given by
\begin{equation}
	m_{\rm fin}= \left\{
                     \begin{array}{ll}
                           0.58+0.22(m_{\rm ini}-1), 
	                           & \quad \mbox{for $m_{\rm ini} 
                                     < 4.7\, [\msun]$}, \\
                           0,  
                                   & \quad \mbox{for $4.7 \le 
                                     m_{\rm ini} \le 8.0 [\msun]$}, \\
                           1.4,    & \quad \mbox{for $8.0 < 
                                     m_{\rm ini} \le 15.0\, [\msun]$}.
                     \end{array}\right.
\end{equation}
Stars with mass $m_{\rm ini} \lesssim 0.46\,\msun$ do not lose any
mass in a stellar wind as they turn into white dwarfs.

The implementation in the \FP\, model is as follows: The mass of the
$k$-th component (see \S\,\ref{FPmassfunction}) is decreased from its
initial mass $m_k^{\rm ini}$ to the final (remnant) mass $m_k^{\rm
fin}$ linearly with time $t$ during the interval $t_{\rm
MS}(m_{k+1/2}^{\rm ini}) < t < t_{\rm MS}(m_{k-1/2}^{\rm ini})$, where
$t_{\rm MS}(m)$ is the main-sequence lifetime of a star with mass $m$.

\section{Initial Conditions}\label{sec:initial}

The initial conditions for our Fokker-Planck and \nbody\ calculations
are identical to those of the survey performed by CW.  It is assumed
that the galaxy has a flat rotation curve and the star clusters are on
circular orbits 
with $v_g = 220$\,km/s 
around the galactic center.
The tidal radius changes with time as $r_{\rm t}(t) \propto
M^{1/3}(t)$.  The initial density profiles of the star clusters are
set up from King models (King 1966).  Calculations are performed with
three different values of $W_\circ =$ 1, 3 and 7, where $W_\circ$ is
the dimensionless central potential of the King models.  The initial
mass function is given by
\begin{equation}
dN \propto m^{-\alpha}\,dm,
\end{equation}
with $\alpha=$1.5, 2.5 and 3.5 between $m = 0.4~\msun$ and $15~\msun$.

In addition to the two dimensionless parameters $W_\circ$ and
$\alpha$, there is one more free parameter: the initial relaxation
time.  Following CW, the mean relaxation time within the tidal radius
is defined as
\begin{equation}
	t_{\rm rlx,CW} = \frac{M^{1/2} r_{\rm t}^{3/2}}
                              {G^{1/2} m_\star  \ln N},
\end{equation}
(eq. [6] of CW with $r=r_{\rm t}$ and $c_1=1$), where $m_\star$ is
a representative stellar mass. This equation can be rewritten with setting
$m_\star =\msun$ as
\begin{equation}
	t_{\rm rlx,CW} = 2.57\times 10^6\, \mbox{yr} \; F_{\rm CW},
        \label{eq:trcw}
\end{equation}
where
\begin{equation}
	F_{\rm CW} \equiv \frac{M}{[\msun]}\frac{R_{\rm g}}{\rm [kpc]}
                 \frac{\rm [220~km\,s^{-1}]}{v_{\rm g}}
                 \frac{1}{\ln N} \label{eq:family}
\label{eq:fcw}
\end{equation}
defines the cluster family (see CW).  Here $R_{\rm g}$ is the distance
to the Galactic center and $v_{\rm g}$ is the circular speed of the
cluster around the Galactic center (cf.\, eqs. [1] and [2] of CW).  The
crossing time of the cluster is then
\begin{equation}
	t_{\rm cr} \approx \left(\frac{r_{\rm t}^3}{GM} \right)^{1/2}
                   =   t_{\rm rlx,CW} \frac{[\msun]}{M} 
                          \ln N . \label{eq:tdyn}
\end{equation}

A group of clusters of the same family (the same $F_{\rm CW}$)
have the same initial relaxation time (\ref{eq:trcw}).  Our survey covers
CW's families 1, 2, 3 and 4. Table\, \ref{tab:fam} reviews the
properties of the selected families.  For convenience Table\,
\ref{tab:fam} also gives the more usual half-mass relaxation time
(Spitzer 1987, p.\ 40),
\begin{eqnarray}
	t_{\rm rh} &=& 0.138\frac{M^{1/2}r_{\rm h}^{3/2}}
                               {G^{1/2} \langle m \rangle \ln N} 
				\nonumber\\
	           &=& 3.54 \times 10^5\, \mbox{yr}\; F_{\rm CW}
                            \left(\frac{r_{\rm h}}{r_{\rm t}}
                            \right)^{3/2}
                            \frac{[\msun]}{ \langle m \rangle },
\end{eqnarray}
where $r_{\rm h}$ is the half-mass radius and $\langle m \rangle$ is
the mean stellar mass.

\placetable{tab:fam}


\section{Comparison between Fokker-Planck and $N$-body Models}
\label{sec:FPvsNB}

We compare the results of anisotropic \FP\ calculations (\Aa\ models)
and those of \nbody\ calculations.  The escaping stars in the \FP\
models are removed on a crossing time scale using Eq.~(\ref{eq:LO}),
but first we have to calibrate the value for \aesc.

\subsection{The calibration of \aesc}\label{sec:aesc}

The rate of mass loss in the \FP\ models changes monotonically with
\aesc; larger \aesc\ causes the models to lose mass on a shorter time
scale.  Increasing \aesc\ has qualitatively the same effect on the
mass loss speed as increasing $N$ (i.e. increasing $t_{\rm rlx}/t_{\rm
cr}$) with \aesc\ held fixed.

The right choice of the initial conditions with which
\aesc\ should be calibrated is rather subtle (TPZ).
We use a wide range of initial
conditions to find the most appropriate value for \aesc.  No
statistical analysis is performed for deciding which value of \aesc\
is the best, since
the criterion with which a statistical test should comply is
not clear.
Rather we judge from global agreement in mass evolution by eye.

\begin{figure}
\psfig{file=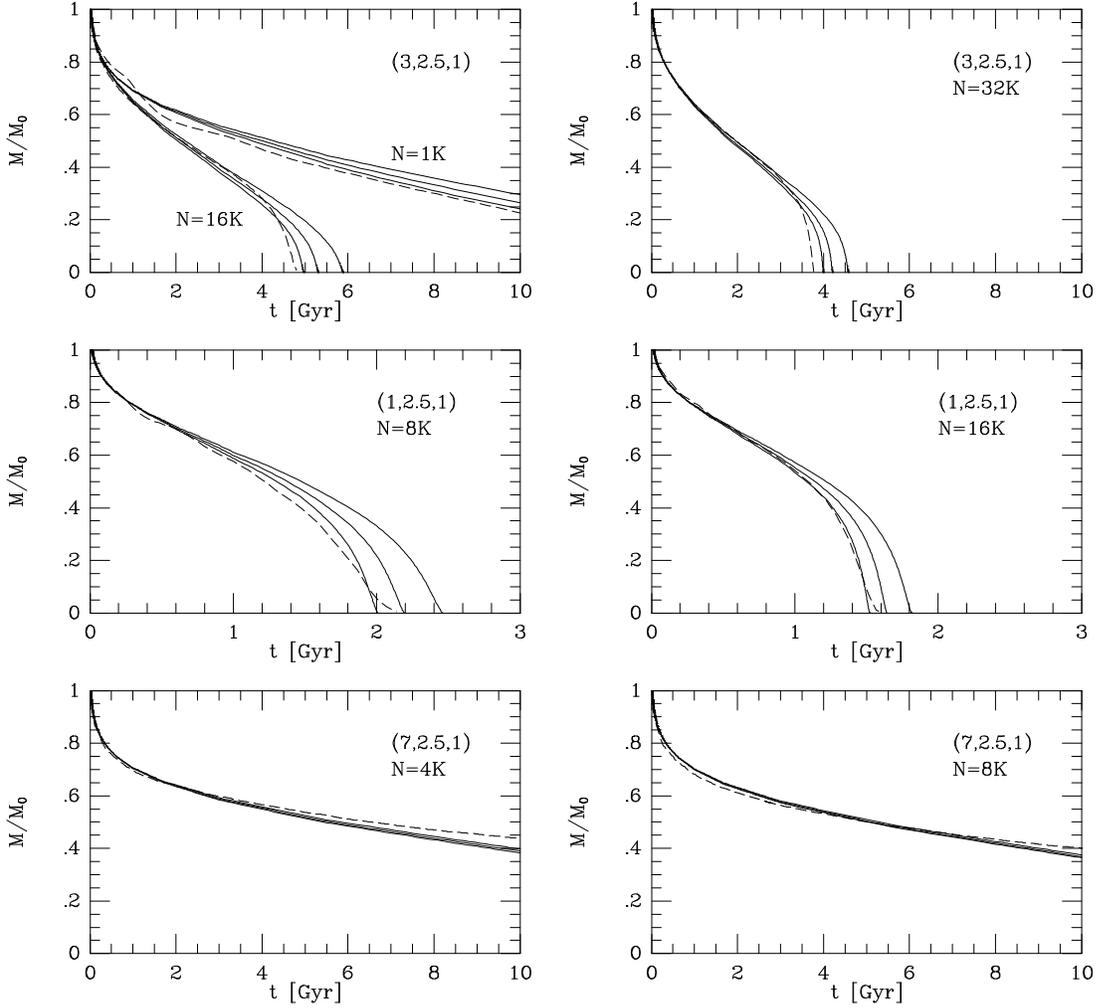,width=15cm}
\figcaption[aesc.ps] 
{Comparison of \nbody\ models (dashed
lines) with Aa \FP\ models with $\aesc=$2, 2.5, and 3 (the
lower, middle, and upper solid lines, respectively) for the evolution
of the total mass.
Larger \aesc\ causes faster mass loss.  
Top panels: 
the initial conditions are
$(W_\circ,\alpha,\mbox{family})=(3,2.5,1)$.
The average of 10 runs is shown for the $N=$1K \nbody\ model.
Middle panels:
$(W_\circ,\alpha,\mbox{family})=(1,2.5,1)$.
Bottom panels:
$(W_\circ,\alpha,\mbox{family})=(7,2.5,1)$.
\label{fig:aesc}}
\end{figure}

\placefigure{fig:aesc}

Figure \ref{fig:aesc} shows the
total mass evolution for \FP\ calculations with $\aesc=$2, 2.5, and 3
for selected initial models and compares them with \nbody\
calculations.  These initial models were selected because stellar
evolution as well as two-body relaxation play a significant role for
these models.

For the models shown in the top two panels,
$\aesc=3$ gives good
agreement between the \FP\ and \nbody\ models at later epochs for $N=$1K;
For $N=$16K and 32K, a value of \aesc\ between 2 and
2.5 is preferable, except for the final phase.  
For the models shown in the middle panels,
$\aesc=3$ gives good agreement for the entire evolution.
For the models shown in the bottom panels,
the evolution is  
rather insensitive to \aesc\ as well as $N$, and the agreement between
the \FP\ and \nbody\ models is good.

From these comparisons we conclude that the best agreement between \nbody\ and
\FP\ models is achieved for $2 < \aesc < 3$, and we adopt the
median value of $\aesc = 2.5$ for the rest of our calculations.

\subsection{The time evolution of the total mass}\label{sec:comp-mass}

Here we present comparisons between \nbody\ and \FP\ calculations
for various initial conditions.
A part of the results were already presented in the previous
subsection, but are reproduced here with new information.

\placefigure{fig:fp-nb}

TPZ performed \FP\ calculations for the initial model of
$W_\circ=3$, $\alpha=2.5$ and family 1 with
using $\aesc=2$.
Fig.~\ref{fig:fp-nb}a shows
repeated calculations with the new adopted value of $\aesc=2.5$.
We also perform \nbody\ and \FP\ calculations with initially 8K and
16K stars for other selected sets of initial conditions (for economical
reasons one set of initial conditions is computed with only 4K and 8K
stars); see Figs.~\ref{fig:fp-nb}b--f.
The \nbody\ calculations are continued up to
10 Gyr unless disruption occurs before this time.  To see the
convergence for $N \rightarrow \infty$ one extra \FP\ calculation for
each set of initial conditions is performed with the instantaneous
escape condition.

The dash-dotted lines in Fig.~\ref{fig:fp-nb}
represent the expected evolution of the total mass
when no escapers are allowed, i.e., when mass loss occurs only
through stellar evolution.
They give upper limits to the cluster mass at each instance.  
(The \nbody\ model of Fig.~\ref{fig:fp-nb}a
with 1K stars seems to lose mass less quickly than the pure
stellar-evolution mass loss case! But this is caused by random
rendering in the small number of stars.)

\placefigure{fig:fp-err}

\placefigure{fig:xi}

Fig.~\ref{fig:fp-err} shows the time evolution of the relative
difference in the total mass between the \FP\ and \nbody\ models,
which we define as $[M_{\mbox{F-P}}(t)-M_{\nbody}(t)]/M_{\nbody}(t)$.
The rapid increase of the relative difference seen in the upper two panels
of Fig.~\ref{fig:fp-err} is the result of the disruption of the
\nbody\ models.

Fig.~\ref{fig:xi} plots the dimensionless mass loss rate $\xi$
as a function of time, where time is expressed in units of the initial
half-mass relaxation time $t_{\rm rh0}$.
The cluster mass loss rate $dM/dt$ may be divided into two parts; 
\begin{equation}
\frac{dM}{dt} = \left( \frac{dM}{dt} \right)_{\rm esc}
              + \left( \frac{dM}{dt} \right)_{\rm se} ,
\end{equation}
where the first term in the right-hand side represents the mass loss
due to escapers from the tidal radius and the second term represents
the mass loss due to stellar evolution.
Since the stellar evolution is the same in the \FP\ and \nbody\
models, it may be useful to separate the stellar evolution mass loss
from the total mass loss in order to further clarify the difference
between the two models.
Thus in Fig.~\ref{fig:xi} we plot the mass loss rate due to
escapers $\xi_{\rm esc}$ as well as the total mass loss rate
$\xi_{\rm tot}$ defined as follows:
\begin{equation}
\xi_{\rm tot} = - \frac{t_{\rm rh0}}{M} \frac{dM}{dt}, \;\;
\xi_{\rm esc} = - \frac{t_{\rm rh0}}{M}
\left(\frac{dM}{dt}\right)_{\rm esc} .
\end{equation}
When there is no stellar evolution, $\xi_{\rm esc}$ is usually much
less than 1 (Lee \& Goodman 1995). This is to be expected,
since in that case mass loss proceeds only on two-body relaxation time
scale.
In the present models, since the stellar evolution mass loss causes
the shrink of the tidal radius and thus produces more escapers,
$\xi_{\rm esc}$ exceeds 1 sometimes.

Fig.~\ref{fig:fp-nb} as well as Figs.~\ref{fig:fp-err} and \ref{fig:xi}
show that
the \FP\ and \nbody\ calculations agree fairly well over the
entire range of initial conditions we investigated.
Apparently the single value of $\aesc = 2.5$ is applicable.
The relative difference of the total mass at each time
is generally less than 10\%. However, for the cases of 
$(W_\circ,\alpha, \family)=(3,2.5,4)$ and $(7,1.5,4)$,
the difference is rather large.
This large difference seems to well correlate with
large $\xi_{\rm esc}$ ($>1$), especially in the \nbody\ models.
Also when the cluster is about to dissolve (see panels [a] and [b]),
$\xi_{\rm esc}$ becomes larger than 1 and the \FP\ and \nbody\ models deviate.

In the case of $(W_\circ,\alpha,\mbox{family})=(3,2.5,4)$, the
\nbody\ models tend to lose mass more quickly than the \FP\
models.  The mass loss in the \nbody\ models accelerates rapidly,
compared with the \FP\ models, from about 1~Gyr for the 16K model and
from about 2~Gyr for the 8K model, and after some time the rate of
mass loss decreases again.
A similar behavior, with smaller
amplitudes, is also observed for the other initial conditions.  This
phased mass loss is caused by a retardation effect in the \nbody\ models.  
Each time the clusters lose mass by stellar evolution and by
loosing stars from the tidal boundary it requires several crossing times
to fill up the outer
layers of the cluster and to recover dynamical equilibrium. 
In this period mass
loss mainly comes from stellar evolution only, which results in a
temporary decrease in the rate of mass loss.
Since the crossing time is long ($t_{\rm cr} \sim 1$ Gyr) in the
models shown in Fig.~\ref{fig:fp-nb}d,
such a phased mass loss is clearly seen in this case.
(In the \FP\ models dynamical equilibrium is assumed.)
The large values of $\xi_{\rm esc}$ result from 
the rapid stellar-evolution-driven mass loss and the long relaxation
time of these models
($\xi_{\rm esc}$ measures the fraction of mass lost per relaxation time).

The models with $(W_\circ,\alpha,\family)=(7,1.5,4)$ are characterized
by a flat initial mass function and a high initial concentration.
Fig.~\ref{fig:fp-nb}e shows that the cluster loses about 80\%
of its initial mass within the first billion years, and still remains
bound.  The mass loss is mainly driven by
stellar evolution.  The rather flat initial mass function causes about
33\% (in number) of the stars to leave the main-sequence within
1 Gyr. The fraction of mass lost from the cluster due to stellar
evolution alone is therefore about 62\%.  Nevertheless the cluster
still survives beyond 10 Gyr.  The cluster is saved by its initial high
central concentration ($W_\circ=7$). In the first 1 Gyr the halo of the
cluster is stripped almost completely, but the compact core survives.
Even in this extreme case the agreement between \FP\ and \nbody\
models is fair.
The difference between the \FP\ and \nbody\ models generally builds up
in the first billion years (when $\xi_{\rm esc} >1$),
and stays almost constant afterwards.
The \FP\ models assume that the potential change is adiabatic,
but this assumption is clearly violated during the early rapid mass-loss
period, since $t_{\rm cr} \sim 1$ Gyr.

The main results of this subsection are summarized as follows:
1) Aa \FP\ models with $\aesc=2.5$ agree fairly well with
\nbody\ models in all the cases we investigated;
2) however, the agreement becomes worse when a large fraction of
the total mass is lost on a crossing time scale.
Such rapid mass loss conflicts with the assumption
of the adiabatic potential change which the \FP\ model is based on.
In real globular clusters the number of stars is at least one order
of magnitude larger than the number of particles used in the present
\nbody\ simulations,
and therefore we expect that the \FP\ model should work well for them.

\subsection{The mass function}

At the start of each calculation the mass function
is independent of the distance to the cluster center. Mass loss by
stellar evolution and mass segregation due to the dynamical evolution
of the star cluster cause the global mass function as well as the
local mass function to change with time.  After some time the mass
function becomes noticeably a function of the distance to the cluster
center.

Figure \ref{fig:w7a2.5f1.mf} shows the mass function for the \nbody\
and the \FP\ calculations for the model with $W_\circ=7$, $\alpha=2.5$
and family\, 1 at the age of 10\,Gyr.  At this moment the clusters has
lost about 60\% of its initial mass (see
Fig.~\ref{fig:fp-nb}c).  The differences between the global
mass-functions of the \nbody\ model and of the \FP\ model are
negligible.  Even the partial mass functions for the main-sequence
stars and for the compact objects (white dwarfs and neutron stars)
agree excellently.  At an age of 10\,Gyr the mass function for the
main-sequence stars is somewhat flatter than the initial mass
function, but the shape of the mass function is still well represented
by a single power-law.  Also for some other initial models we compared
the mass functions at $t=10$~Gyr between \FP\ and \nbody\ models, and
found good agreement in all cases.

\placefigure{fig:w7a2.5f1.mf}

\section{Fokker-Planck Survey Results}\label{sec:FPsurvey}

In this section we present the results of
our anisotropic \FP\
calculations in which escaping stars are removed instantaneously from
the cluster.  These calculations give lower limits to the lifetimes of
the clusters (see TPZ). 
We use the same stellar evolution model as CW in the survey simulations.

\placetable{tab:sur}

Table \ref{tab:sur} summarizes the survey results (\Aa\ models) and
compares them with the results obtained by CW (their Table 5).
Following CW, the destiny of the cluster is classified as `C' for core
collapse or `D' for disruption.  The mass of the cluster $M_{\rm end}$
(in units of the initial mass) and the time $t_{\rm end}$ (in Gyr) at
the moment we stop the calculations (resulting in C or D) are listed
in Table \ref{tab:sur}.  The listed end time for collapsing clusters
means the moment of core collapse. These clusters
will survive beyond the core collapse,
but their post-collapse evolution is not followed in our calculations.

The disruption time as well as the mass at that instant are rather ill
defined in \FP\ calculations.  By definition a cluster ceases to exist
(disrupts) at the moment $M = 0$.  However \FP\ calculations usually
break earlier.  Mass loss often accelerates rapidly before the moment
the cluster ceases to exist (see Fig.~\ref{fig:surm-a}).  In such a
case the Fokker-Planck model fails to find the self-consistent
solution of the Poisson equation, and breaks (see CW, \S\ VIIa).  CW
identified this moment as disruption time.  This may be associated
with the moment when the cluster loses dynamical equilibrium
(Fukushige \& Heggie 1995).  CW's definition, however, is not free
from ambiguity particularly for the mass at disruption, since the
above-mentioned mass-loss acceleration is related to the
breaking of numerical calculations.

For clusters that dissolve we continued the calculations until the
mass evolution curves start to bend downwards very rapidly (see 
Figure~\ref{fig:surm-a}).  We define $t_{\rm vert}$ as the moment
when the mass loss rate becomes virtually infinite.
Then, however, 
the mass at $t_{\rm vert}$ is not well defined.
Therefore, for convenience,
we decided to take $t_{\rm end}=0.99 t_{\rm vert}$ as the disruption
time, and then $M_{\rm end}=M(t_{\rm end})$.
We confirmed by eye that the point $(t_{\rm end}, M_{\rm end})$
determined by the above definition
corresponds to the ``turn-off point" of the mass evolution curve.

\subsection{The total mass and the concentration as a function of time}

\placefigure{fig:surm-a}

\placefigure{fig:surc-a}

Figure \ref{fig:surm-a} shows the evolution of the total mass for all the
initial models.  The circles plotted at the ends of some of the lines indicate
the moments when those clusters experience core collapse.
The destiny of the clusters is
determined by the initial conditions ($W_\circ$, $\alpha$, and
$F_{\rm CW}$).  The early evolution is strongly affected by the choice of
$W_\circ$ and $\alpha$, and in less extent by the initial relaxation
time, simply because the latter requires more time to affect the
dynamics of the stellar system.  The lifetimes of the clusters which
are disrupted within $\sim 1$\,Gyr, for example, are almost
independent of the initial relaxation time.  The evolution of these
clusters is mainly driven by stellar mass loss. This causes the
clusters to expand after which the Galaxy gobbles up the
clusters.  Core collapse occurs only in some long-lived ($>3$~Gyr)
clusters.  For these clusters the mass loss rate gradually decreases
with time (see also Portegies Zwart et al.\ 1998).

Figure \ref{fig:surc-a} shows the evolution of the concentration
parameter $c$, which we define by (cf. King 1966)
\begin{equation}
c \equiv \log_{10} \frac{r_{\rm t}}{r_{\rm c}} .
\label{eq:c}
\end{equation}
Here the core radius $r_{\rm c}$ is
\begin{equation}
r_{\rm c} \equiv \left(\frac{3v_{\rm m0}^2}{4\pi G\rho_0}\right)^{1/2} ,
\label{eq:rc}
\end{equation}
where $\rho_0$ is the central density and $v_{\rm m0}$ is the
density-weighted velocity dispersion in the cluster center.  Note
that this core radius $r_{\rm c}$ and the King core radius (King 1966) 
are not
identical even for single-mass systems, since the central velocity
dispersion is not used in the definition of the King radius.  The two
quantities get closer for more concentrated (higher $W_\circ$)
clusters (cf.\ Binney \& Tremaine 1987, p.~235).

In some cases (for $W_\circ=7$, $\alpha=1.5$ of all the families, and for
$W_\circ=1$, $\alpha=3.5$ of family 1) the cluster experiences core
collapse after almost all ($\sim 99$\%) of the initial mass is lost.
We identify these clusters as collapsing clusters, because the
concentration $c$ rapidly increases at the final epochs (see
Fig.~\ref{fig:surc-a}).  If the initial mass of a cluster is
relatively small (e.g.\, $M_0 \lesssim 10^4~\msun$), the definite
evidence of core collapse will be very hard to observe, since the number of
stars in the core will become too small (e.g. $N_{\rm core}\lesssim
10$). Deep core collapse in such systems can easily be prevented by
the formation of binaries by three-body processes or by burning
primordial binaries. Such processes are more effective in small $N$
clusters.

For $W_\circ=7$, $\alpha=1.5$, irrespective of the families, the
clusters have very small masses (compared to their initial masses) around
$t=$10~Gyr.  This implies that they may be observed at present as tiny
globular clusters or as old open clusters (see \S~\ref{sec:obs}).

For all the models with $\alpha=1.5$, mass loss due to stellar
evolution causes the concentration to decrease with time before core
collapse occurs. For $\alpha=$2.5 and 3.5, the concentration does not
decrease so much but stays almost constant.  If, at any time, the
concentration becomes smaller than $c \approx 0.4$ (compare with
$c=0.58$ for the $W_0=1$ King model), the cluster quickly dissolves.
As CW pointed out (see their section VIIc), the (scaled) density
profiles just before disruption are always similar regardless of
initial models.

\subsection{Comparison with Chernoff \& Weinberg (1990)}\label{sec:ccw}


\subsubsection{Clusters that disrupt}

All our clusters of model \Aa\, live longer than those of CW for the
same initial conditions.  
The largest (relative) discrepancy is about a factor of
ten which is found for $(W_\circ,\alpha)=(1,2.5)$ and
$(3,2.5)$.  For the models with $(W_\circ,\alpha)=(1,1.5)$ and
$(3,1.5)$, the discrepancy is about a factor of two.  In the
latter models the mass functions are very flat and the potentials
are shallow, and thus stellar mass loss alone is enough to disrupt these
clusters.  It is therefore not surprising that even the \Aa\ models
cannot live so long compared with the CW models.
The disruption time hardly depends on the initial relaxation time
for those clusters which
disrupt within $\sim1$\% of the half-mass relaxation time.

\placetable{tab:sur-f1}

Table \ref{tab:sur-f1} compares our \Ae\ and \Aa\ models with CW's
models for family 1.  In addition to the data already given in Table
\ref{tab:sur}, $t_{\rm end}$ and $M_{\rm end}$, two new quantities,
$t(M_{\rm end,CW})$ and $M(t_{\rm end,CW})$, are presented.  Here
$M_{\rm end,CW}$ and $t_{\rm end,CW}$ denote the final mass and time,
respectively, of each of CW's models which are given in Table
\ref{tab:sur}.  These two quantities are particularly useful for
comparing different models for disrupted clusters.


The energy criterion for the tidal boundary is used for the CW models
as well as for the \Ae\ models.  The \Ae\
models are in this sense in between the CW and \Aa\ models.
Comparison with the \Ae\ models helps us to understand how the apocenter criterion
affects the results.  The \Ae\ models lose mass faster than the \Aa\
models in all the cases shown in Table \ref{tab:sur-f1}, simply
because the energy criterion introduces a larger escape region in
phase space than the apocenter criterion (Takahashi et al.\ 1997; TPZ).
The difference between the CW and \Ae\ models is rather small in all
cases.  For each set of initial conditions the two models have the
same destiny, disruption or collapse.  The difference in the
disruption time is within a factor of two, and that in $t(M_{\rm
end,CW})$ is even smaller. We therefore conclude that the
introduction of the apocenter criterion plays the most important role
in extending the cluster lifetime.

%

For $(W_\circ,\alpha)=(1,2.5)$ and $(3,2.5)$ we find large differences
between the CW and \Aa\ models not only in $t_{\rm end}$ and $M_{\rm end}$ 
but also in $t(M_{\rm end, CW})$ and $t(t_{\rm end,CW})$.  This assures that
the discrepancy between the two models are not merely due to
differences in the definitions of the disruption time and mass.  The
clusters survive beyond the disruption times given by CW, which is
also supported by \nbody\ simulations (TPZ; \S\ \ref{sec:FPvsNB}).

\subsubsection{Clusters that collapse}

About half (seventeen) of the clusters calculated with \Aa\ models
experience core collapse, while in the calculations of CW less than
30\% (ten) of the clusters experience core collapse.  This means that
seven of the models of CW are disrupted while the corresponding \Aa\
models experience core collapse.  In these \Aa\ models core collapse
occurs much later than the disruption time of the CW models; a too
high rate of mass loss in the CW models disrupts the clusters too
quickly.

All the clusters which experience core collapse according to CW do
experience core collapse according to our calculations also.  Core collapse
occurs after two-body relaxation starts to dominate the evolution of
the clusters, i.e.; when stellar-evolution mass loss has sufficiently
decreased. Since
CW's models generally overestimate the rate of mass loss, all models that
finally develop a collapsed core according to their calculations
experience core collapse in our calculations too.

For $(W_\circ, \alpha) = (7,2.5)$ and $(7,3.5)$ of all the families,
both the models of CW
and our \Aa\ models experience core collapse.  
For these models, the moment of core collapse
approximately scales with the initial relaxation time.
This indicates that their evolution is
dominated by two-body relaxation, not by stellar evolution.  
In those cases, the differences in the moment of core
collapse between CW and \Aa\ models are less than 10\,\%.

The introduction of anisotropy into the models itself does not have very significant
effects on the moment of core collapse (see Takahashi 1995; Takahashi
et al.\ 1997).  Core collapse proceeds almost independently of the
outer parts of the cluster and the anisotropy in the core is always
small.  The difference between the energy and apocenter criteria
affects mass loss from the tidal boundary, and the mass loss affects core
collapse time by reducing the relaxation time.  But the effect of
the mass loss on core collapse time
is only significant for low-concentration clusters which
lose a substantial fraction of their mass before collapse (cf. Quinlan
1996).

In addition, when mass loss proceeds slowly, on the relaxation time
scale, the difference between the apocenter criterion and the energy
criterion is less important than when mass loss proceeds more rapidly
driven by stellar evolution.
The stars which exit in a region lying between
the boundary of the energy
criterion and that of the apocenter criterion in phase space
(see, Takahashi et al.\ 1997, Fig.~1; or TPZ, Fig.~1)
are responsible for slower mass loss in apocenter-criterion models.
These stars diffuse over the tidal boundary 
on a time scale of the order of the relaxation time.
If mass loss by stellar evolution is ineffective and the cluster
evolves slowly on the relaxation time scale
the difference between apocenter- and energy-criterion models
will be small (cf. Takahashi et al.\ 1997). On the other hand,
if the cluster loses mass rapidly due to stellar evolution
compared with the relaxation time scale,
the above-mentioned phase-space region will
play an important role as a mass reservoir.

We therefore conclude that the moment of core collapse is correctly
calculated with isotropic \FP\ models as well as with anisotropic \FP\
models as long as the cluster is relaxation-dominated.

%

\placefigure{fig:sur}

Figure \ref{fig:sur} gives a graphical representation of
Table\,\ref{tab:sur} for family 1.  The largest discrepancies occur
around a boundary which separates disrupted clusters and
collapsing clusters; the disrupted clusters are on the lower left
region and the collapsing clusters are on the upper right region in
this figure.

For $W_0=3$ and $\alpha=3.5$ of family 1 and 2, 
the differences in the collapse time
between isotropic and anisotropic models are rather large.  These models are
near the separating boundary mentioned above;
the CW models of family 1 and 2 collapse but those of family
3 and 4 dissolve.  In such cases, even a small difference in the mass
loss rate may cause a large difference in the fate of the cluster.

\subsection{Fokker-Planck Survey Results for $N=3\times10^5$}\label{sec:sur3e5}

Real star clusters contain a finite number of stars,
typically $N \sim 10^5$ -- $10^6$. It is not clear
whether this number is large enough to be treated as infinite,
though it is usually assumed in \FP\, calculations.
For example, it is not clear whether the instantaneous escape condition,
which is valid for $N \rightarrow \infty$ and is adopted in our survey
simulations, is a good approximation for typical globular clusters.
To test the effect of a finite crossing-time,
we performed a series of calculations by using the crossing-time
removal condition, Eq.~(\ref{eq:LO}), with $N=3\times10^5$.

\placefigure{fig:surm-a.3e5.f4}

Figure \ref{fig:surm-a.3e5.f4} presents the mass evolution for models
of family 4 with $N=3\times10^5$ initially.  Family 4 has the largest
relaxation time and thus has the largest crossing time for fixed $N$.
The initial total masses of these models are $7.3\times10^5~\msun$,
$3.0\times10^5~\msun$, and $2.0\times10^5~\msun$, and the crossing
times $t_{\rm cr}$ are 0.03~Gyr, 0.06~Gyr, and 0.10~Gyr, for
$\alpha=1.5$, 2.5, and 3.5, respectively.  These models are compared
with the models with the instantaneous escape condition (see
Fig.~\ref{fig:surm-a}).

The relative difference between the models with $N=3\times10^5$ and
those with $N \rightarrow \infty$ for the clusters which dissolve
within a few Gyr is rather large.  In these cases the models with
$N=3\times10^5$ live a factor of two or three longer than the $N
\rightarrow \infty$ models.  The lifetime of these clusters is
comparable to the crossing time.  The evolution of these models is
therefore very sensitive to the way escapers are removed from the
clusters; crossing-time removal or instantaneous removal.  For
long-lived clusters that survive for a Hubble time the details of how
escaping stars are removed become less important; the instantaneous
removal condition gives a good approximation for these cases.

%

\section{Discussion}\label{sec:discussion}

\subsection{The anisotropic Fokker-Planck model}

We have shown that
our anisotropic \FP\ (\Aa) models are in excellent agreement with
\nbody\ models for a wide range of initial conditions.  Although
expected from theoretical considerations, it is still quite
striking that a single value of \aesc\ suffices to achieve agreement
between the very different types of numerical models and over such a
wide range of initial conditions.  Subtle differences between the
Fokker-Planck and \nbody\ results, however, remain.  These are
probably due to the greater detail and higher accuracy of the \nbody\
models compared to the \FP\ models. (The absence of binary formation
via three-body processes in the \FP\, models may also play a role
here.)  In some cases, the relatively simplistic implementation of the
tidal field via a cut-off radius causes a somewhat peculiar behavior
in the \nbody\ models (see \S~\ref{sec:TFNbody}).  \FP\, models lose
their credibility when the time scale for mass loss becomes comparable
to the dynamical time scale, since in that case the assumption of an
adiabatic change in the potential is violated.

Concerning the models shown in the present paper,
all the \Aa\ models live longer
than the isotropic models.
However,
there are cases where anisotropic
models live longer than isotropic models.  For example,
Takahashi et al.\ (1997)
show the results that, in the absence of stellar evolution,
\Aa\ and \Ae\ models evaporate faster than an isotropic
model in the tidal field.  
The shorter lifetime of these anisotropic models is induced by
two-body relaxation at the inner regions which causes a quick emergence of
elongated-orbit stars in the halo.  The apocenters of these
stars are frequently outside the tidal radius.  Especially when the
tidal field is relatively weak mass loss in anisotropic models is
quicker than in isotropic models (see Takahashi \& Lee 1999 for details).

The apocenter criterion is considered more realistic than the energy
criterion, but also this approach has its limitation.  The
implementation of a spherical cutoff radius instead of a
self-consistent tidal field, for example, limits the conditions on
which escapers are identified.  However, also in a real cluster a star
must pass the tidal radius in order to leave the cluster.  The
apocenter criterion is therefore an important improvement over the
simple energy criterion used in previous \FP\, calculations.  The
implementation of the apocenter criterion, however, requires
anisotropic \FP\ models; isotropic \FP\ models do not contain
sufficient information about the stars' orbits to apply the apocenter
criterion.  Unfortunately it seems hard to adjust the energy criterion
used in isotropic models
such that the mass loss rate becomes similar to that for the apocenter
criterion.  This is because the apocenter criterion causes the
cluster's mass-loss rate to depend on the angular-momentum
distribution (anisotropy) of the stars near the tidal boundary, and
the anisotropy behaves in a rather complex way: it changes with time,
and depends on the mass of the stars and the distance to the
cluster center (Takahashi 1997; Takahashi et al.\ 1997).

%

\subsection{Effect of the tidal field}\label{sec:TFNbody}

To investigate the effect of the self-consistent tidal field compared
with the adopted tidal cut-off, we perform several \nbody\
calculations with different implementations of the tidal field. The
same initial conditions as for the models in Figure \ref{fig:fp-nb}a are
adopted.
For each model, calculations are performed ten times using
different sets of random numbers
with 1024 stars.  The results are
presented in Figure \ref{fig:w3a2.5f1.1k.mass}.

\placefigure{fig:w3a2.5f1.1k.mass}

The solid line in Figure \ref{fig:w3a2.5f1.1k.mass} represents the
mean of ten calculations with a simple tidal cut-off (model TCR1,
the $\sigma/2$ deviations from the mean are shown in
Figure \ref{fig:fp-nb}a).  The other lines represent the
results for the calculations with a self-consistent tidal field.  For
all the tidal-field models the same tidal force is used but different
cut-off (maximum) radii are selected.
Stars in the tidal-field models are removed as soon as
they go beyond \rt\ (model TFR1; the T stands for tidal, F for field
and the R stands for the radius, in units of the tidal radius, at
which stars are removed from the simulation), $2\,\rt$ (model TFR2),
$10\,\rt$ (model TFR10), or $100\,\rt$ (model TFR100) from the cluster
center.
For computational reasons it
is convenient to set a cut-off radius even for the calculations where
a tidal field is used.
The mass given in Figure \ref{fig:w3a2.5f1.1k.mass} is the
mass contained in stars which are bound to the cluster.

All the models presented in Figure \ref{fig:w3a2.5f1.1k.mass} are
indistinguishable for the first one billion years, where mass loss is
dominated by stellar evolution. After that, model TCR1 and model TFR1
start to lose mass more rapidly than the other models.  This
acceleration in mass loss is related to a too sudden removal of
escaping stars which is caused by the small cut-off radius.  Model
TFR1 keeps losing mass at a rate slightly higher than the other
tidal-field models.  After about 2 Gyr the mass loss rate of model TCR1
becomes smaller than for the models with a tidal field.  Not
surprisingly, stars are removed more quickly if a tidal field is
incorporated and stars are removed at $r_{\rm t}$.  Figure
\ref{fig:w3a2.5f1.1k.mass} shows that an artificial cut-off radius in
excess of 2\rt\ hardly affects the results in the models where a
self-consistent tidal field is incorporated.

%

The Fokker-Planck model with 1K stars (see
Figure \ref{fig:fp-nb}a) loses more mass in the first billion
years than the \nbody\ models (TCR1, TFR1 to TFR100). This difference may be
caused by the assumed adiabatic response of the \FP\ models to changes
in the potential (Cohn 1979).  The cluster loses about 30\% of
its initial mass within 1~Gyr.  Since the initial half-mass crossing
time is about 0.5~Gyr, this change in the potential is quite
impulsive.  In spite of this breakdown of one of the fundamental
assumptions in \FP\, models, the agreement with the \nbody\ model is,
except model TFR1, fairly good.

We have demonstrated that \nbody\ and \FP\ models with the tidal
cut-off behave in a similar fashion as \nbody\ models with the
self-consistent tidal field, for one selected set of initial conditions.
However, it is not obvious whether tidal cut-off models are always
similar to tidal-field models.  Giersz \& Heggie (1997) also
investigated the difference between these two kinds of \nbody\ models.
They concluded that the mass evolution is not much affected by the
difference between the two implementations of the tidal boundary
conditions.  We like to encourage further investigations on this
issue.

\subsection{Comparison with similar \nbody\ surveys by other authors}

Recently Aarseth \& Heggie (1998) investigated the time scaling of
$N$-body models with respect to the number of stars.  They
incorporated the Galactic tidal field in the same way as Fukushige \& Heggie
(1995) did.  The same stellar evolution tracks as used by CW were adopted.

Aarseth \& Heggie (1998) introduced ``variable time scaling'' which
couples the stellar evolution time scale to the time scale on which
the star cluster evolves dynamically.  During the early stages of the
evolution of the cluster, when the effect of stellar evolution is dominant,
time scaling based on the crossing time is applied.
At later stages, when two-body relaxation becomes dominant, the
relaxation time is used for time scaling.  
This variable time scaling has a qualitative
physical basis, but there remains uncertainty in when and how to
change from one scaling to another.

Using the variable time scaling Aarseth \& Heggie (1998) find good
agreement between their \nbody\ calculations and the \FP\ calculations
of CW for those clusters that survive long enough for their evolution
to become dominated by two-body relaxation.  For quickly disrupted
clusters, the models of CW live too short compared to the models of
Aarseth \& Heggie (1998).  These conclusions are qualitatively
consistent with our conclusions on the comparison between our \FP\
models and the CW models.  A detailed comparison between Aarseth \&
Heggie's (1998) $N$-body models and our \FP\ models is complicated by
the difference in the implementation of the tidal field.

\placetable{tab:fhah}

In Table \ref{tab:fhah} our \Aa\ models are compared with the \nbody\
results of Aarseth \& Heggie (1998, Table 3) as well as
those of Fukushige \& Heggie (1995, Table 3) for family 1, in terms of core
collapse time and mass, and disruption time and mass.  The models of
the former authors are denoted by ``AH", and those of the latter by
``FH".  A number associated with each of these model names indicates
the galactocentric distance $R_{\rm g}$ (in kpc) at which that model
is placed.  The initial cluster mass is $5.45\times 10^4 \msun$ at
$\sim 10$\,kpc, and $1.49\times 10^5 \msun$ at $\sim 4$\,kpc.  The \Aa\
models use the instantaneous escape condition and their evolution is,
within one family, independent of $R_{\rm g}$
Fukushige \& Heggie (1995) use fixed time scaling based on the
crossing time, which is considered to be appropriate for the early evolution of
clusters.  Table \ref{tab:fhah} shows only those models for which they
obtain definite lifetimes.

The disruption time and mass, $t_{\rm dis}$ and $M_{\rm dis}$,
are listed in Table \ref{tab:fhah}.
The \nbody\, models disrupt at the moment the mass has dropped
to zero, but, as we discussed above, the \FP\, 
calculations usually stop well before the mass goes to zero.
Therefore we should be aware that
$t_{\rm dis}$ for the \nbody\ models and that for \FP\ models
do not have exactly the same meaning.

For $(W_0,\alpha)=(1,2.5)$ and $(3,2.5)$ the results of CW are very
different from those of FH and AH, but our \Aa\, models are consistent with
the results of FH and AH.  For models with the flat initial mass
function of $(W_0,\alpha)=(1,1.5)$ and $(3,1.5)$, the results of CW
deviate largely from the calculations of FH and AH.  
For these cases, the lifetimes of
our \Aa\ models ($\sim$0.02~Gyr) are about twice as long as those
of CW's models, but they are still shorter than the lifetimes of
the models of FH and AH by a
factor of a few.  This difference will be reduced when the crossing-time
removal condition is applied for \FP\ models.
As discussed in \S\, \ref{sec:sur3e5},
for such short-lived clusters with small $N$, the
instantaneous removal condition gives somewhat too short lifetimes (see
Fig.~\ref{fig:surm-a.3e5.f4}).

The \Aa\, models (as well as CW's models)
with $(W_0,\alpha)=(7,2.5)$ and $(7,3.5)$ experience
core collapse at about the same time and at about the same cluster mass as
the models of AH.  For $(W_0,\alpha)=(1,3.5)$ and $(7,1.5)$
the \Aa\ models collapse while the \nbody\ models do not.  The
masses of these \Aa\ models at the moment of core collapse, however, are only
$\sim 0.01 M_0$, and thus it is not surprising that the core collapse does
not occur in the \nbody\ systems comprising initially of only a few thousand
particles.

\subsection{Comparison with the observations}\label{sec:obs}

For a better comparison, all our calculations are performed with the
same initial conditions as adopted by CW.  These initial conditions,
however, are probably not the best choice for real globular clusters.
Recent observations have revealed that mass functions are ill
represented by a single power-law (e.g., Scalo 1986; Kroupa et al.\ 
1990)\nocite{sca86}\nocite{1990MNRAS.244...76K} and that the lower
stellar mass limit should be $\lesssim 0.1\msun$ (e.g, De Marchi \&
Paresce 1995a; 1995b; Marconi et al.\ 1998; see also Chernoff 1993).
\nocite{1998MNRAS.293..479M}
\nocite{1995A&A...304..211D}\nocite{1995A&A...304..202D} Another
uncertainty, which may seriously affect the simulation results, is the
initial compactness of star clusters.  The King models from which we
selected the initial density profiles may not be the best choice for
the initial density profiles for tidally-truncated star clusters
(Heggie \& Ramamani 1995).\nocite{1995MNRAS.272..317H} It is not even
clear whether real globular clusters were born filling up their tidal
lobes. Initial under-filling of the tidal lobe
may extend the cluster lifetime considerably.
Regardless of these uncertainties in the initial conditions,
it will be still worthwhile
to discuss some implications for the observed population of
globular clusters.  

Our calculations indicate that clusters which are 
observed in a state of core collapse at present must have been born fairly
concentrated, otherwise core collapse would not be occurring within the
age of the Universe. At core collapse most of the models have
lost $\apgt 50$\% of their initial mass.  This indicates that observed
globular clusters in a state of core collapse were born with a high
concentration and considerably more massive than today's mass.

%

All the clusters with a concentration parameter $c \approx 1$ at $t
\approx 10$\,Gyr appear to be in a local minimum in
concentration. These clusters recover their concentrations later as
core collapse follows.  A cluster which reaches a concentration of $c
\approx 0.4$ at any time evaporates quickly thereafter.  The least
concentrated globular clusters in our Galaxy have a King concentration
parameter of $c_{\rm King} \approx 0.5$ (Trager et al.\ 1993).  This is
consistent with our findings.\footnote{Note, however, that $c_{\rm
King}$ is not identical to the $c$ (see \S~\ref{sec:FPsurvey}) and
that the one-$\sigma$ error in the observed $c_{\rm King}$ is about
0.2 (Trager et al.\ 1993).}

%

According to our survey
most of globular clusters with a rather flat initial mass function
($\alpha=1.5$) disrupt within a Hubble time. 
This indicates that the majority of the
present-day clusters probably had steeper initial mass functions.
However, we cannot exclude the possibility that some of the globular clusters
were born with an initial mass function of $\alpha \sim 1.5$ 
but have still survived owing to
the initial high concentration.
For example, the model with $W_\circ=7$, $\alpha=1.5$ and $\family\, 3$
loses about 99\% of its initial mass and has still a high concentration
at 12~Gyr (see Figs.~\ref{fig:surm-a} and \ref{fig:surc-a}).  The old
open cluster Berkeley 17 has similar characteristics as this model, 
as we show below.

Phelps (1997) derives an age for Berkeley 17 of 10\,Gyr to 13\,Gyr
(but Carraro et al., 1999, estimates an age of $9 \pm 1$ Gyr), a distance
to the Galactic center of $R_{\rm g}=11$~kpc, a size of $\sim 5$~pc,
and a total mass in visible stars of about $400\,\msun$.

The core of the model with $W_\circ=7$, $\alpha=1.5$ and family 3
collapses at an age of 12\,Gyr. At this moment the total mass is only
1.3\% of its initial mass (see Table~\ref{tab:sur}).  Substitution of
$R_{\rm g}=11$~kpc and $v_{\rm g}=220$~km/s into
Eq.~(\ref{eq:family}) results in an initial mass of $M_0 \approx
2.3\times10^5~\msun$.  At the moment of core collapse the cluster has
then a mass of $\sim 3000~\msun$ and the half-mass radius of $\sim
4$~pc.  The majority of this mass is hidden in stellar remnants, i.e.:
$\sim 900 \msun$ in neutron stars and $\sim 1700 \msun$ in white
dwarfs, and only $\sim 400 \msun$ is in main-sequence stars
(see also Heggie \& Hut 1996; Vesperini \& Heggie 1997 for dark remnants
in globular clusters).  This
model cluster looks similar to Berkeley 17 in the size and luminous mass.


We note that
the fraction of neutron stars in our models is overestimated, since many are
likely to escape from the cluster by a ``kick'' at the time of a
supernova explosion. An intrinsic asymmetry in the supernova is
believed to give neutron stars a kick velocity which is generally
much higher than the escape velocity from the cluster (Drukier 1996;
Portegies Zwart \& van den Heuvel 1999). This effect is not taken into
account in our calculations.

%

\placefigure{fig:w7a1.5f3.mf}

Figure \ref{fig:w7a1.5f3.mf} shows the global mass function for the
model with $W_\circ=7$, $\alpha=1.5$, and family 3, at $t=12$\,Gyr.
The filled circles and the open circles represent the mass functions
for the white dwarfs and for the main-sequence stars, respectively.
The mass function for the main-sequence stars peaks at $m \sim 0.8
\msun$.  
The stars with $m \apgt 0.8 \msun$ have evolved off the main-sequence.
The peculiar shape of the
main-sequence mass function for $m \aplt 0.8 \msun$, which decreases
with decreasing $m$, is the result of the preferential loss of low
mass stars through the tidal boundary; high mass stars are more
centrally concentrated than low mass stars due to mass segregation and
therefore have less chance to escape. In general the global mass
function becomes flatter as the cluster loses mass (e.g.\, Vesperini
\& Heggie 1997).  However, it is rare to find such an inverted global
mass function as is shown in Fig.~\ref{fig:w7a1.5f3.mf} in our
surveyed models.  Such a mass function only appears in clusters which
have lost $\apgt 99\%$ of their initial mass on the relaxation time
scale.  These ``remnant'' clusters were formerly the cores of more
massive clusters. High mass stars are over-abundant in such cores as a
result of mass segregation.

%

%

The mass function of the globular cluster NGC 6712 around the half-light
radius also peaks near $0.75\,\msun$ (De Marchi et al.\ 1999), and
looks similar to the mass function shown in
Fig.~\ref{fig:w7a1.5f3.mf}. 
(Although Fig.~\ref{fig:w7a1.5f3.mf} shows the global
mass function, the local mass function at the half-mass radius
is similar to that.)

De Marchi et al.\ (1999) interpreted the observed inverted mass
function as evidence for severe tidal disruption,  which is
consistent with the above model.
Thus we suspect that NGC
6712 was born far more massive than what it is today.  
With the current mass of $\sim 10^5 \msun$ (Pryor \& Meylan 1993), its
initial mass would be $\sim 10^7 \msun$, if that model is applied.
However, this prediction of the initial mass is rather uncertain
quantitatively, because our survey
is limited in many respects.
The range of initial conditions covered by our survey
is still limited, as mentioned above.
Further we assume a steady tidal field.
The orbit of NGC 6712 is, in fact, likely to be very eccentric;
the cluster is presently at $R_{\rm g}=3.5$ kpc,
but the minimum and maximum galactocentric distances
are predicted to be about 0.3 kpc and 7 kpc, respectively 
(Dauphole et al.\ 1996).
Therefore, bulge and disk shocks
(see Gnedin \& Ostriker 1997; Murali \& Weinberg 1997;
and references therein) have probably affected the mass evolution of
this cluster considerably, as pointed out by De Marchi et al.\ (1999).

%

Since we argued above that 
Berkeley 17 may be represented by
the model shown in Fig.~\ref{fig:w7a1.5f3.mf},
it will be interesting to observe the mass function of Berkeley 17
down to very low mass stars.

\section{Conclusions}\label{conclusions}

We have investigated the evolution of star clusters including
the effects of the
tidal field of the parent galaxy and mass loss from stellar evolution
by using {\it anisotropic} \FP\ models as well as \nbody\
models.  Our new \FP\ models, which include a
new implementation of the tidal field, agree much better with
\nbody\, models than {\it isotropic} \FP\, models.

The new implementation of the tidal boundary consists of two parts:
the apocenter criterion
assures that only those stars are removed from the stellar system for
which apocenter radius exceeds the tidal radius of the cluster;  the
crossing-time removal condition prevents escaping stars from being removed
in an infinitesimal time step, but keeps them in the stellar system
for some time of the order of the crossing-time.  The crossing-time removal
introduces an additional free parameter,  \aesc, to the \FP\, model.
We find that a single value of $\aesc = 2.5$ provides good agreement
between \nbody\ and \FP\ calculations over a wide range of initial
conditions.  These improvements to the \FP\, model allow us to
perform calculations in much wider ranges of parameter space,
including varying the number of stars.  This opens the possibility to
compare \FP\ calculations directly with direct \nbody\ calculations.

With the improved \FP\, model we have studied the evolution of globular
clusters over a wide range of initial conditions.  The surveyed
initial parameters are identical to those adopted by CW.  Our clusters
live generally longer than the isotropic models of CW.
The largest differences appear in cases 
where clusters dissolve promptly in
the tidal field before core collapse occurs,
except for cases of too prompt dissolution due to too severe
mass loss via stellar evolution.
For some
initial conditions, our models reach core collapse, while the models of
CW dissolve.  
The differences are
rather small for long-lived clusters
which start with steep mass functions and
high concentrations and finally reach core collapse. 
These findings are consistent with the conclusions of Aarseth \&
Heggie (1998) who compare their \nbody\ models with the models of CW.

About half of the clusters calculated in our survey are disrupted
within 10\,Gyr. 
Only four of our 36 samples
experience core collapse within 10~Gyr, and two of the four
have lost more than 99\% of their mass at
the moment of core collapse.  
This suggests that the range of the initial conditions
with which clusters reach core collapse within the age of
the Universe is rather small in our survey.  
On the other hand,
about 20\% of the known
Galactic globular clusters are classified as post-collapse clusters
(Djorgovski \& King 1986; Chernoff \& Djorgovski 1989; Trager et
al.~1993).  
These post-collapse clusters
might have been born with initial conditions that are preferable
for core collapse at $\sim 10$ Gyr, e.g., $\alpha=2.5$ and $3 < W_\circ
\lesssim 7$. 
However it should be noted that 
not all of possible initial conditions
of real globular clusters are covered by our (and CW's) survey.
In particular the adopted initial mass functions are
probably not the best choice.

In the survey
we find a few peculiar models which have
very small masses ($\sim$ 1\% of the initial mass) but are well
concentrated at $\sim$10\,Gyr.  Such clusters happen to be classified as old
open clusters, like Berkeley 17, rather than as globular
clusters, depending on their luminosities, appearances, and
positions relative to the Galactic disk.
The present-day global mass functions of these model clusters are also
peculiar in the sense that they decrease with decreasing
stellar mass.  A similar inverted mass
function was observed in the globular cluster NGC 6712 by De Marchi et
al.\ (1999).  This indicates that NGC 6712 might have lost a very large
fraction of its initial mass, 
if the observed mass function is really close to the global mass function
and does not only reflect a peculiarity of the initial mass function
of the cluster.

\begin{figure}
\psfig{file=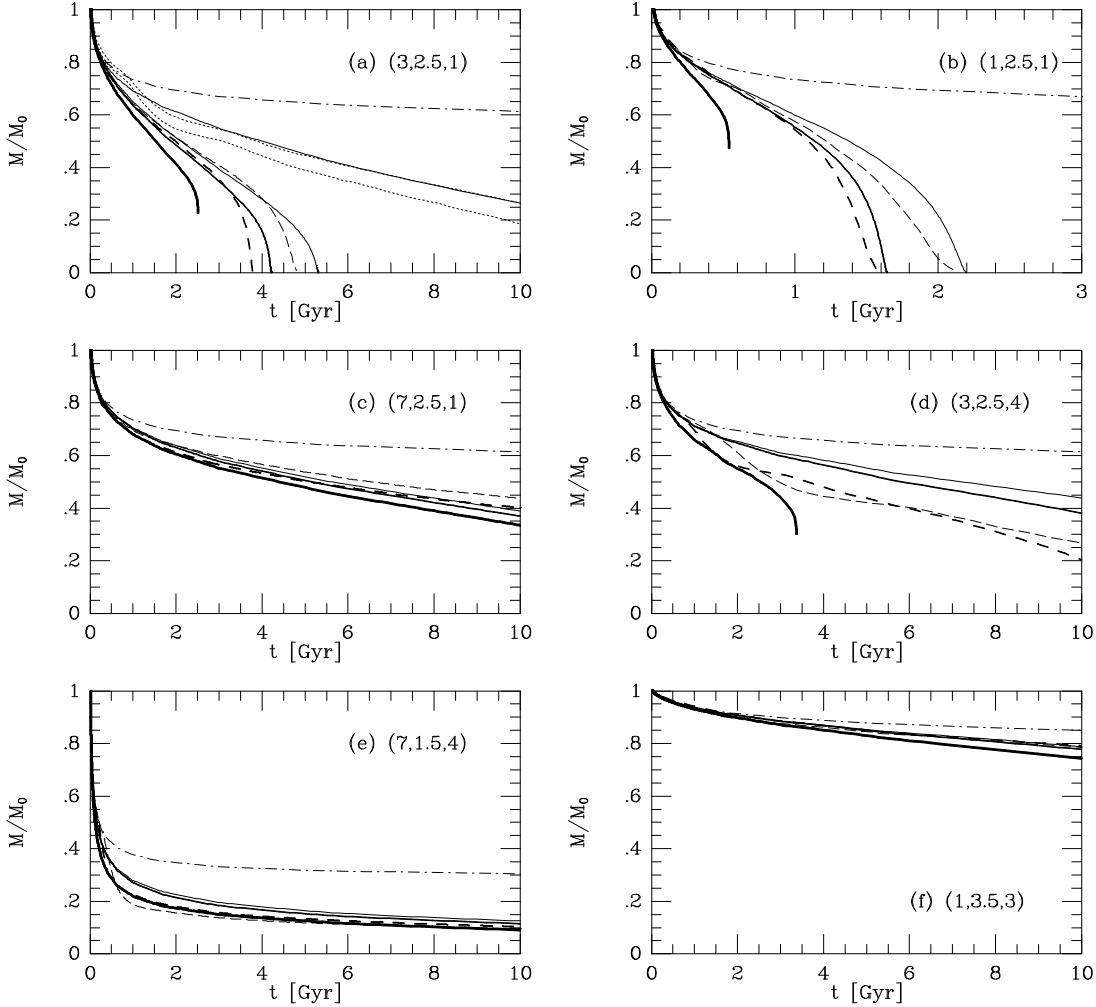,width=15cm}
\figcaption[fp-nb.ps]
{Comparison between Aa Fokker-Planck models (solid lines)
and $N$-body models (dashed lines and dotted lines)
for the evolution of the total mass.
Thickness of the lines stand for largeness of $N$.
The thickest solid line in each panel represents the Fokker-Planck model with
the instantaneous escape condition, which corresponds to the
the limit of $N \rightarrow \infty$.
The other Fokker-Planck models are calculated with the crossing-time
escape condition with $\aesc=2.5$ for finite $N$.
The dash-dotted lines represent the mass evolution expected when
mass loss occurs only through stellar evolution (with no escapers).
(a)
The initial conditions are $(W_\circ, \alpha, \mbox{family})=
(3, 2.5, 1)$;
$N=$1K, 16K, and 32K, from right to left.
Only for the $N=$1K $N$-body model,
$\pm \sigma /2$ deviation from the mean of 10 $N$-body runs
is shown (two dotted-lines).
(b)
$(W_\circ, \alpha, \mbox{family})=(1,2.5,1)$;
$N$=8K, 16K.
(c)
$(W_\circ, \alpha, \mbox{family})=(7,2.5,1)$;
$N$=4K, 8K. 
(d)
$(W_\circ, \alpha, \mbox{family})=(3,2.5,4)$;
$N$=8K, 16K.
(e)
$(W_\circ, \alpha, \mbox{family})=(7,1.5,4)$;
$N$=8K, 16K.
(f)
$(W_\circ, \alpha, \mbox{family})=(1,3.5,3)$;
$N$=8K, 16K.
\label{fig:fp-nb}}
\end{figure}

\begin{figure}
\psfig{file=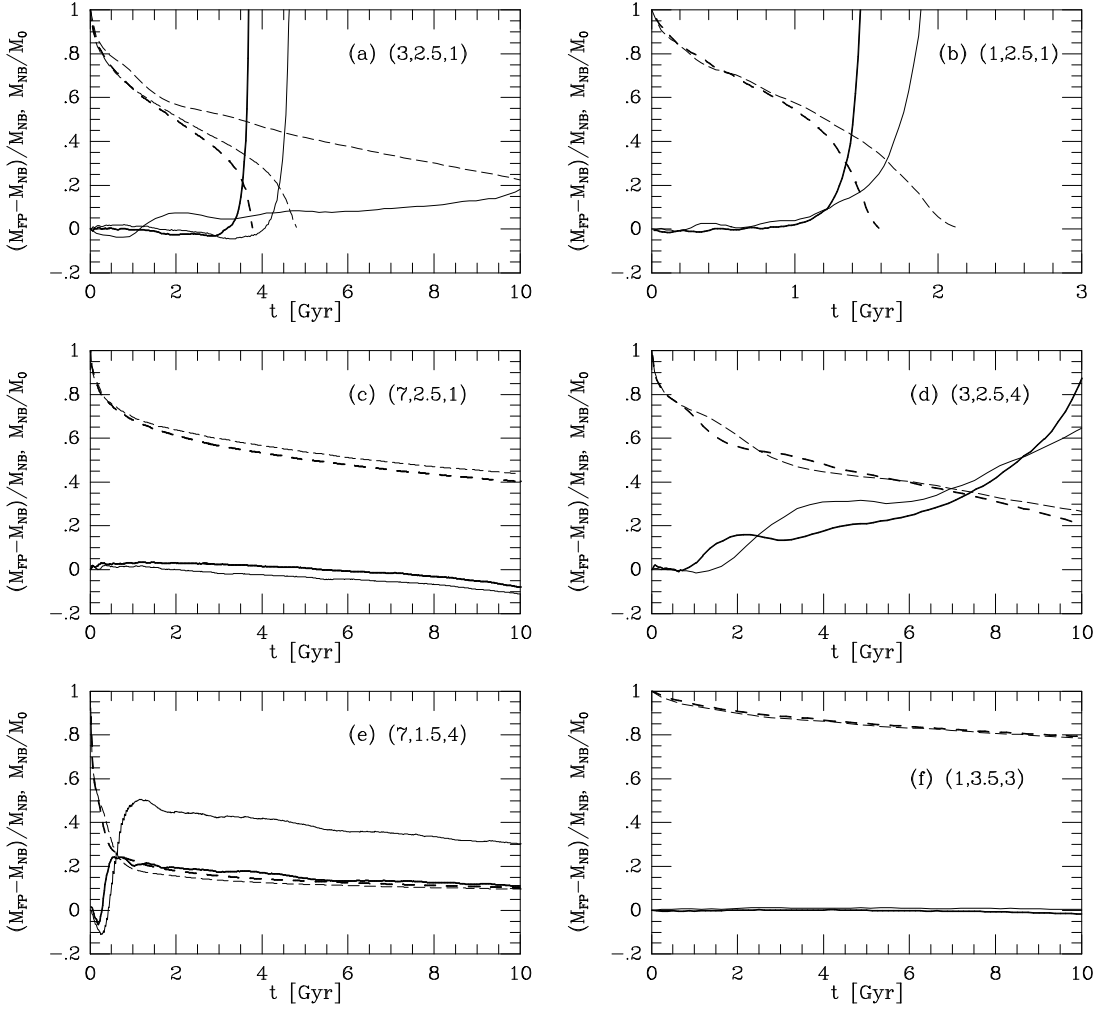,width=15cm}
\figcaption[fp-err.ps]
{Time evolution of the relative mass difference between 
\FP\ and \nbody\ models,
$[M_{\mbox{F-P}}(t)-M_{\nbody}(t)]/M_{\nbody}(t)$ (solid lines),
for the models shown in Fig.~\ref{fig:fp-nb}.
The total mass of the \nbody\ models is also plotted (dashed lines).
In each panel the initial conditions are denoted by three parameters
$(W_\circ, \alpha, \mbox{family})$.
The thick lines represent the largest-$N$ models in each panel.
\label{fig:fp-err}}
\end{figure}

\begin{figure}
\psfig{file=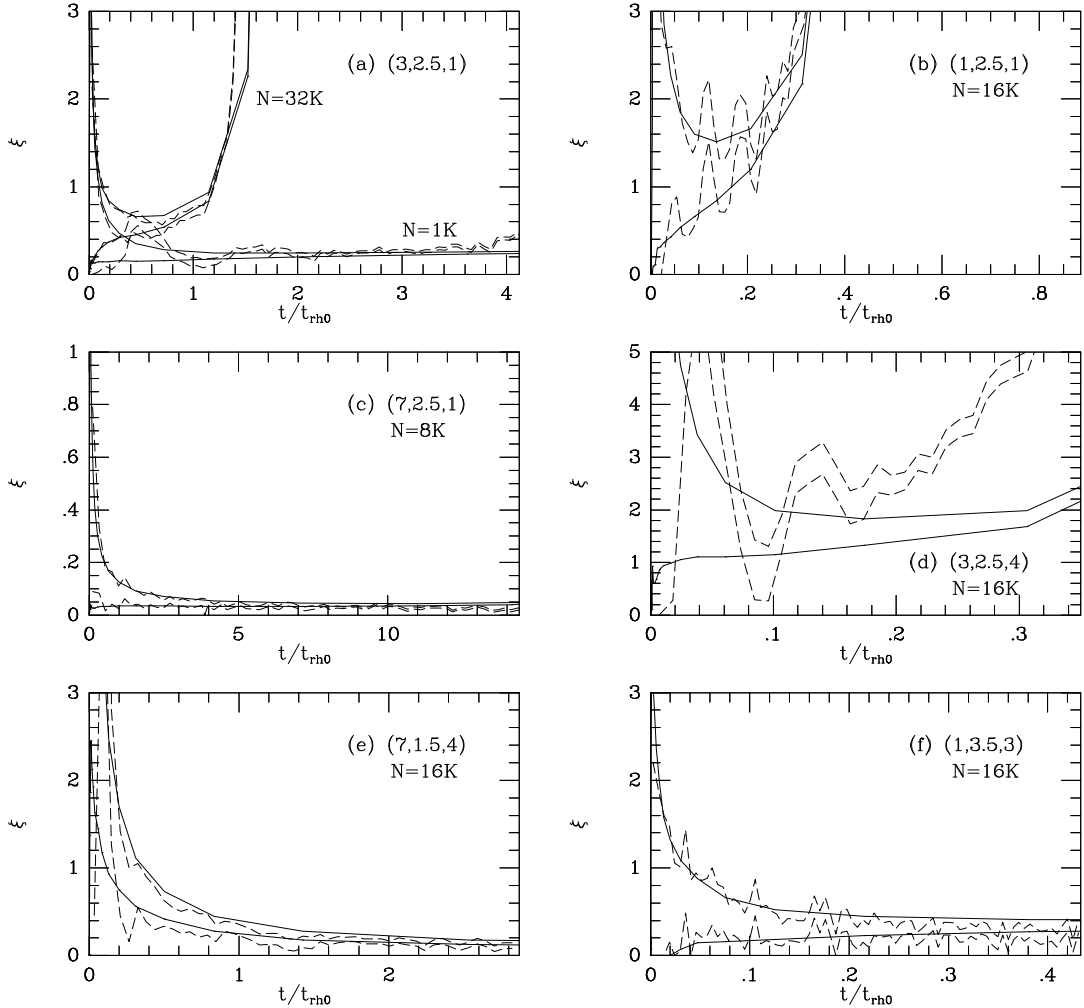,width=15cm}
\figcaption[xi.ps]
{Time evolution of the mass loss rates,
$\xi_{\rm tot}$ and $\xi_{\rm esc}$ ($\xi_{\rm tot} > \xi_{\rm esc}$),
for the Fokker-Planck (solid lines) 
and $N$-body (dashed lines) models shown in Fig. \ref{fig:fp-nb}.
In these plots time is expressed in units of the initial half-mass relaxation
time, but for each panel the plotted period is the same as that in
Fig.~\ref{fig:fp-nb} (i.e., 3 Gyr for [b] and 10 Gyr for the others).
Only the largest-$N$ models are shown for each set of 
$(W_\circ, \alpha, \mbox{family})$,
but in (a) $N=$1K models are also shown.
\label{fig:xi}}
\end{figure}

\begin{figure}
\psfig{file=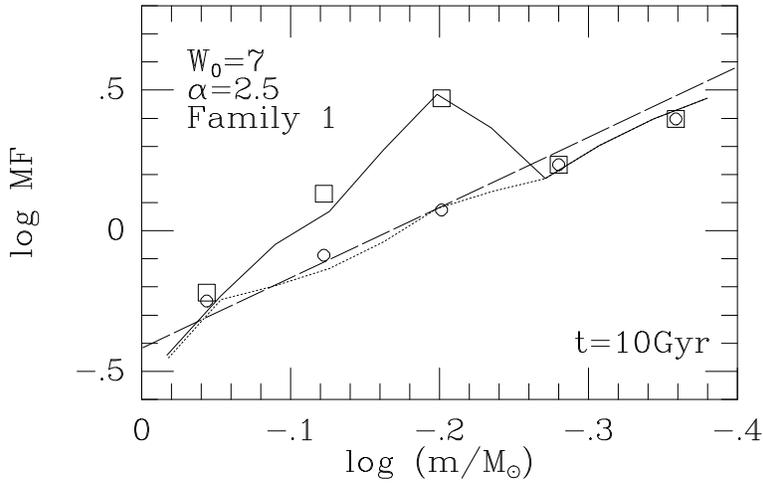,width=11cm}
\figcaption[w7a2.5f1.mf.ps]
{The global mass function for the model with the initial conditions of 
$W_\circ = 7$, $\alpha = 2.5$, and
family\,1 at $t=10$~Gyr.
The dashed line represents the initial mass function.
The mass functions are normalized at each epoch.
The Fokker-Planck results
are given by symbols and the \nbody\ results are given by
lines.
The squares and the solid lines represent the mass functions for all the stars,
and the
circles and the dotted lines represent those for the main-sequence stars only.
\label{fig:w7a2.5f1.mf}}
\end{figure}

\begin{figure}
\psfig{file=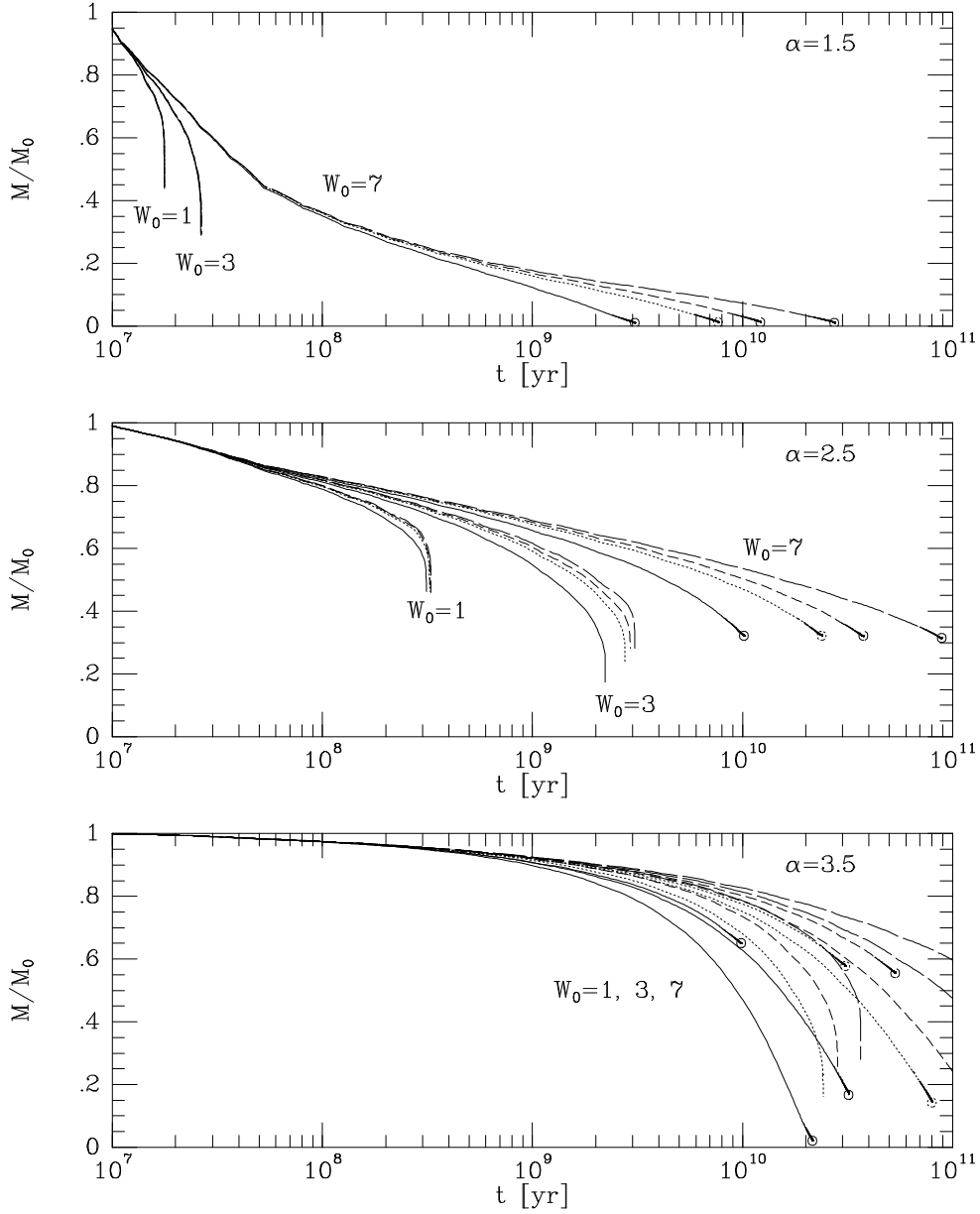,width=15cm}
\figcaption[surm-a.ps]
{Results of the Fokker-Planck survey simulations: Evolution of the total mass.
Three panels from top to bottom show results for the initial mass
functions of $\alpha$=1.5, 2.5, and 3.5.
Families 1 to 4 are plotted with solid, dotted, short dashed, and 
long dashed lines, respectively.
The circles at the end points of some lines indicate that those simulations
stop at the time of core collapse.
\label{fig:surm-a}}
\end{figure}

\begin{figure}
\psfig{file=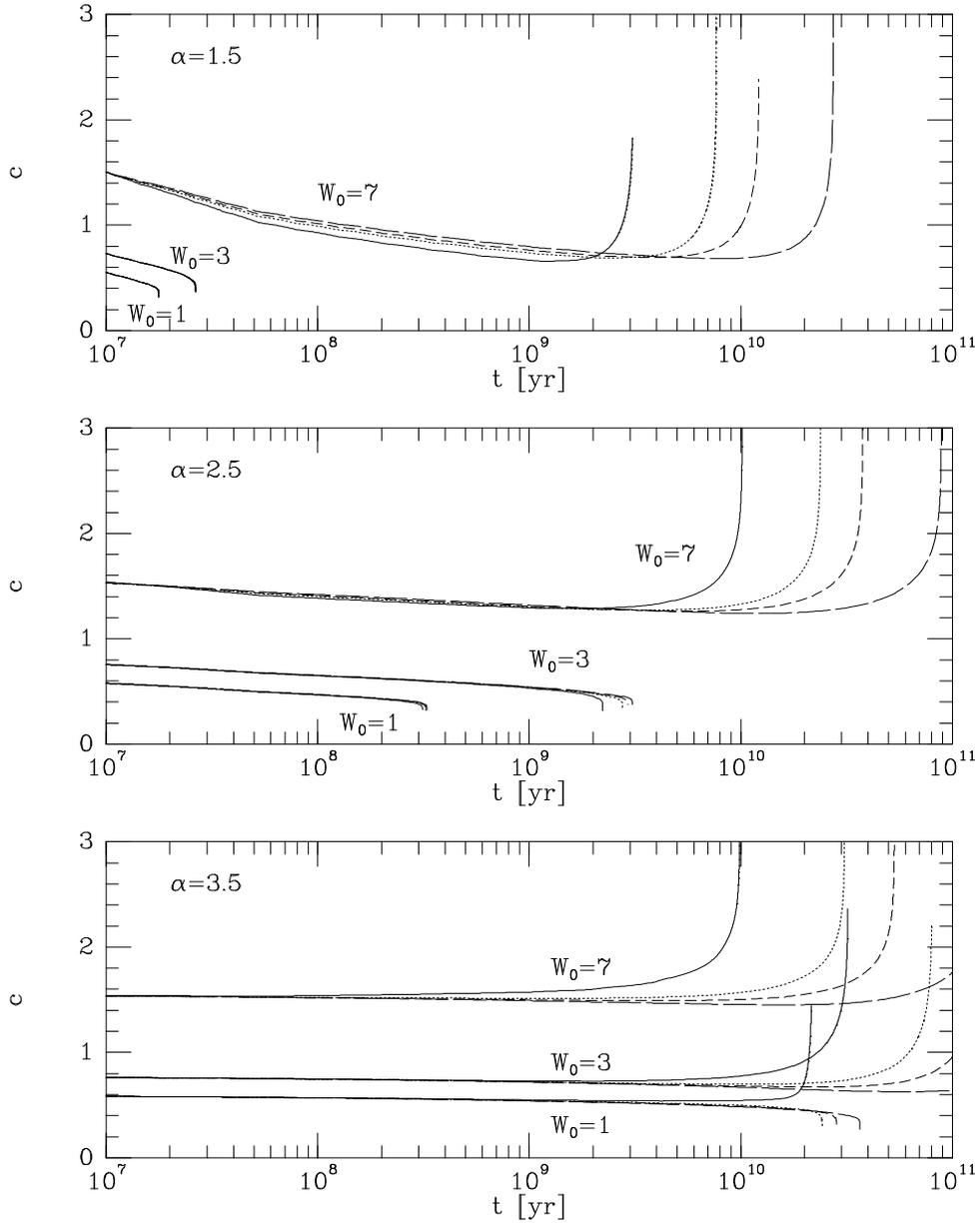,width=15cm}
\figcaption[surc-a.ps]
{Results of the Fokker-Planck survey simulations: 
Evolution of the concentration $c$.
The notation in this figure is the same as that in Fig.~\ref{fig:surm-a}.
\label{fig:surc-a}}
\end{figure}

\begin{figure}
\psfig{file=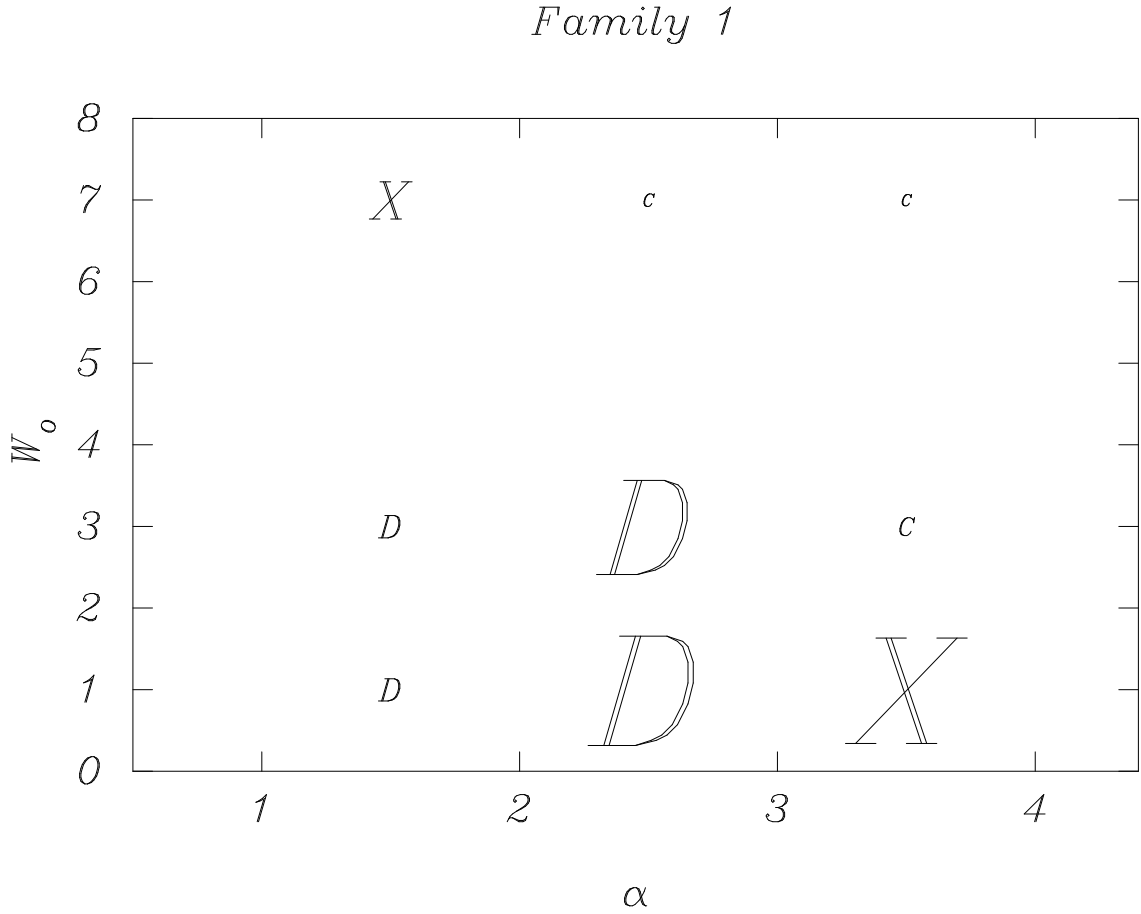,width=15cm,angle=-90}
\figcaption[TPZvsCW.ps]
{Graphical representation of the difference between the Aa models and
the CW models
in the survey results (see
Table \ref{tab:sur}) for family 1.
The power-law index $\alpha$ of the initial mass function
is given along the X-axis and the
initial value for $W_\circ$ is given along the Y-axis.
The symbols $C$ and $D$ indicate the end states of the simulations,
collapse and disruption, respectively. An $X$ indicates that the result
of CW is disruption where the Aa model reaches core collapse.
The size of the letters is proportional to the lifetime at which we
arrive as fraction of the lifetime of the cluster calculated by
CW;
a larger symbol indicates a larger discrepancy. 
\label{fig:sur}}
\end{figure}

\begin{figure}
\psfig{file=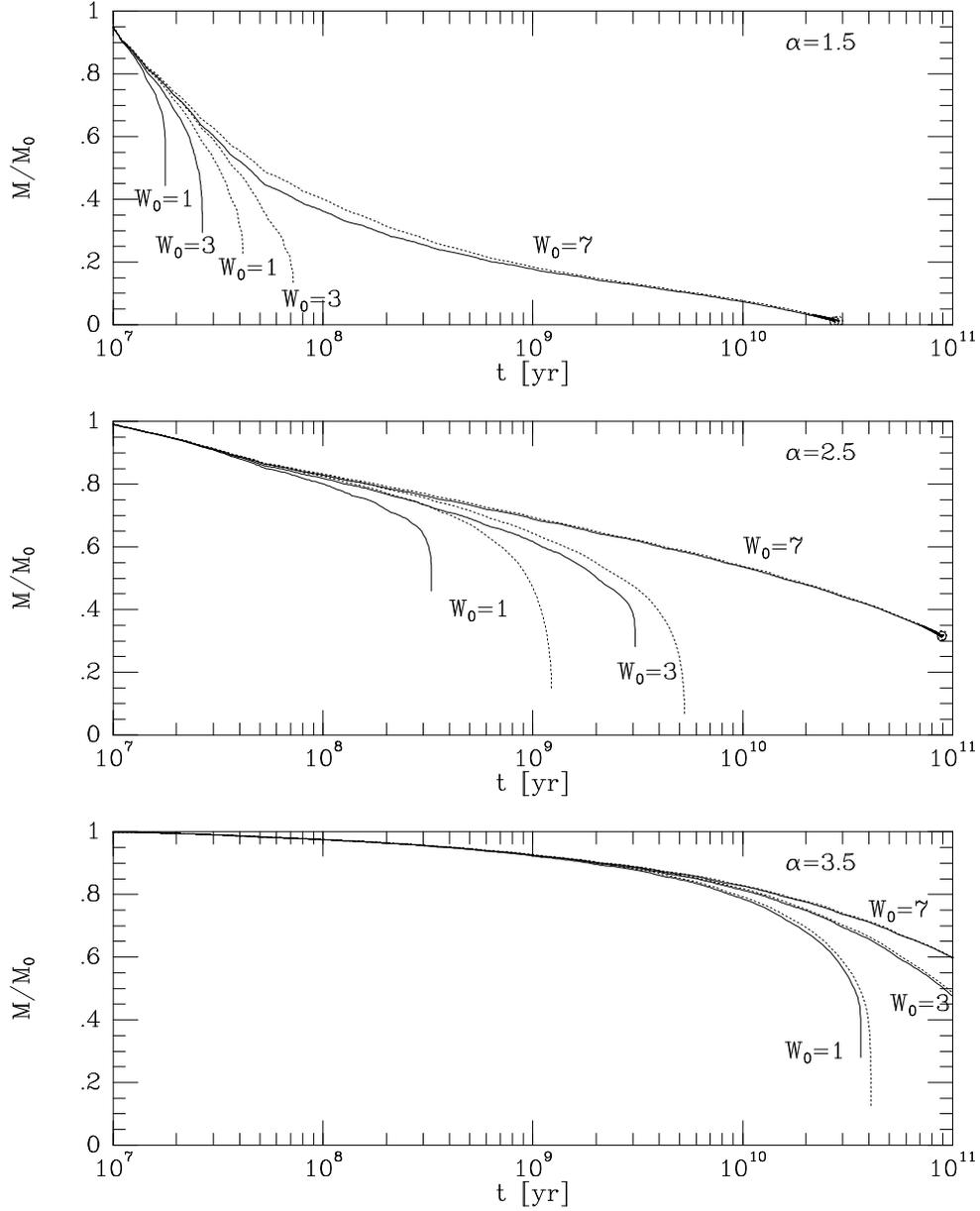,width=15cm}
\figcaption[surm-a.3e5.f4.ps]
{
Comparison of the mass evolution
between the Aa models of $N=3\times10^5$ clusters (dotted lines)
and
the Aa models of $N\rightarrow \infty$ clusters (solid lines,
the models shown in Fig.~\ref{fig:surm-a}).
Only the models of family 4 are shown.
Three panels from top to bottom show results for the initial mass
functions of $\alpha$=1.5, 2.5, and 3.5.
\label{fig:surm-a.3e5.f4}}
\end{figure}

\begin{figure}
\psfig{file=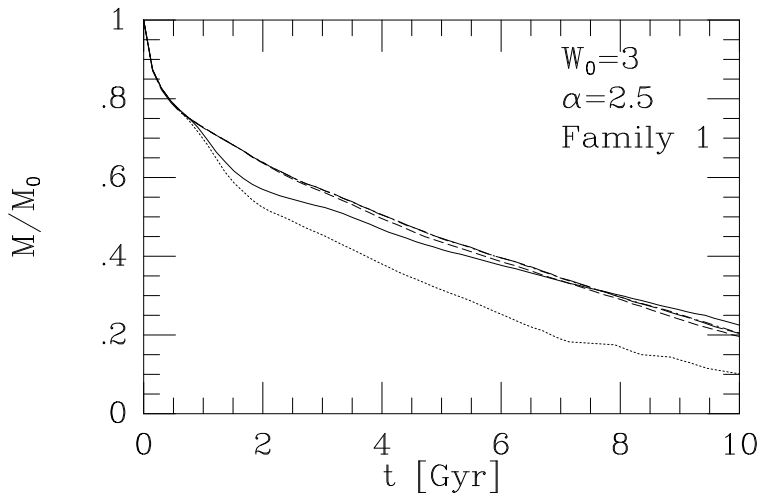,width=11cm}
\figcaption[w3a2.5f1.1k.mass.ps]
{Mass as a function of time
for various \nbody\ models with 1K stars (for $\alpha = 2.5$,
$W_\circ=3$, and family 1) for different implementations of the tidal field.
Each line represents the mean of ten calculations.  The solid line is
calculated with a simple tidal cut-off as is used for all other
\nbody\ calculations in this paper.  The other lines give the results
of calculations with a sef-consistent tidal field in which stars are
removed beyond the radii $\rt$ (dotted), 
$2\rt$ (short dashed), $10\rt$ (long dashed), and $100\rt$
(dash-dotted).  The last two lines are almost indistinguishable.
\label{fig:w3a2.5f1.1k.mass}}
\end{figure}

\begin{figure}
\psfig{file=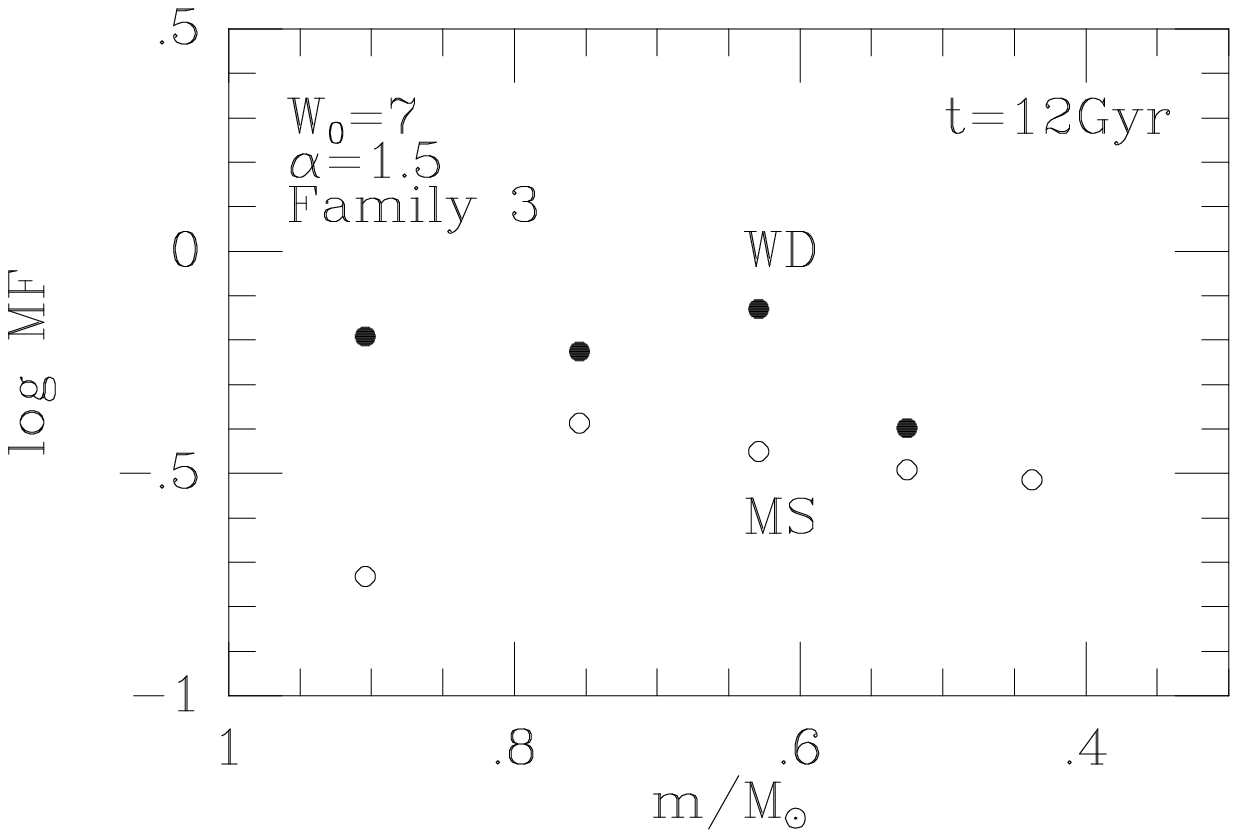,width=11cm}
\figcaption[w7a1.5f3.mf.ps]
{The global mass function for the model with the initial conditions of
$W_\circ = 7$, $\alpha = 1.5$, and family 3, from the \FP\ survey,
at $t=12$~Gyr.
The open circles represent the mass function of the main-sequence stars,
and the filled circles represent that of the white dwarfs.
\label{fig:w7a1.5f3.mf}}
\end{figure}

Clusters which lose $\apgt 50$\% of their initial mass in a very short
time scale may still survive for a long time.  The principle that
losing more than a half of the initial mass of a cluster inevitably
results in disruption is based on the virial theorem and on the
assumption of impulsive mass loss (Hills 1980).  For the initial
conditions of $W_\circ=7$, $\alpha=1.5$, and family 4, more than
half the cluster mass is lost within a few half-mass crossing times
(see Fig.~\ref{fig:fp-nb}e).  But, in fact, 
still the cluster survives for
more than ten billion years and finally experiences core
collapse.

Comparison between \nbody\ calculations with various implementations
of the galactic tidal field and the cut-off boundary condition
indicates that there may be a significant population of stars which
are trapped outside the tidal radius of the cluster but are still
bound to the stellar system.

\acknowledgments We are grateful to Piet Hut, Junichiro Makino, and
Steve McMillan for many discussions and software development. We also
thank Haldan Cohn, Toshiyuki Fukushige, and Douglas Heggie for
enlightening discussions.  SPZ is grateful to University of Tokyo for
its hospitality and for the use of the fabulous GRAPE system.  We wish
to express our special thanks to the anonymous referee for
constructive comments which led us to further careful examinations of
the results.  This work was supported in part by the Research for the
Future Program of Japan Society for the Promotion of Science
(JSPS-RFTP97P01102), and by NASA through Hubble Fellowship grant
HF-01112.01-98A awarded (to SPZ) by the Space Telescope Science
Institute, which is operated by the Association of Universities for
Research in Astronomy, Inc., for NASA under contract NAS\, 5-26555.

\clearpage

\clearpage


\begin{deluxetable}{rrrrr}
\tablecaption{Stellar evolution models. \label{tab:se}}
\tablewidth{0pt}
\tablehead{
\colhead{$m_{\rm ini}/\msun$} & 
\multicolumn{2}{c}{$\log(t_{\rm MS}/{\rm yr})$} &
\multicolumn{2}{c}{$m_{\rm fin}/\msun$} \\
\cline{2-3}\cline{4-5}
& \colhead{CW} & \colhead{SeBa} & \colhead{CW} & \colhead{SeBa}
}
\startdata
0.40 & 11.3 & 11.3 & 0.40 & 0.40 \nl
0.60 & 10.7 & 10.8 & 0.49 & 0.52 \nl
0.80 & 10.2 & 10.4 & 0.54 & 0.56 \nl
1.00 & 9.89 & 10.0 & 0.58 & 0.59 \nl
2.00 & 8.80 & 9.02 & 0.80 & 0.69 \nl
4.00 & 7.95 & 8.29 & 1.24 & 0.80 \nl
8.00 & 7.34 & 7.64 & 0.00 & 1.33 \nl
15.00 & 6.93 & 7.14 & 1.40 & 1.34 \nl
\enddata
\end{deluxetable}

\begin{deluxetable}{lcrrrr}
\tablecaption{Families. \label{tab:fam}}
\tablewidth{0pt}
\tablehead{
\colhead{Family} & \colhead{$F_{\rm CW}$} & \colhead{$t_{\rm rlx,CW}$/Gyr}
& \multicolumn{3}{c}{$(\langle m \rangle/\msun) (t_{\rm rh}$/Gyr)}\\
\cline{4-6}
&&& $W_\circ=1$ & $W_\circ=3$ & $W_\circ=7$
}
\startdata
1 & $5.00 \times 10^4$ & 128 & 3.4 & 2.5 & 0.7 \nl
2 & $1.32 \times 10^5$ & 339 & 9.0 & 6.5 & 1.9 \nl
3 & $2.25 \times 10^5$ & 577 & 15.4 & 11.0 & 3.2 \nl
4 & $5.93 \times 10^5$ & 1522 & 40.6 & 29.1 & 8.3 \nl
\enddata
\tablenotetext{a}{The mean stellar mass $\langle m \rangle=$
2.45\msun, 1.01\msun, and 0.66\msun, for $\alpha=$1.5, 2.5, and 3.5, respectively.}
\end{deluxetable}

\begin{deluxetable}{cccccccccc}
\tablecaption{Survey results. \label{tab:sur}}
\tablewidth{0pt}
\scriptsize
\tablehead{
\colhead{} & \colhead{} & \multicolumn{8}{c}{Family} \\
\cline{3-10}
\colhead{$W_\circ$} & \colhead{$\alpha$} & \multicolumn{2}{c}{1} &
\multicolumn{2}{c}{2} &
\multicolumn{2}{c}{3} & \multicolumn{2}{c}{4}  \\
\cline{3-10}
\colhead{} & \colhead{} & \colhead{CW} & \colhead{Aa} & \colhead{CW} &
\colhead{Aa} & \colhead{CW} & \colhead{Aa} & \colhead{CW} & \colhead{Aa}
}
\startdata
1 & 1.5 & D & D & D & D & D & D & D & D \nl
& & 9.2\sisu{-3} & 0.017 & 9.4\sisu{-3} & 0.018
& 9.3\sisu{-3} & 0.018 & 9.3\sisu{-3} & 0.018 \nl
& & 0.78 & 0.61 & 0.78 & 0.63 & 0.78 & 0.63 & 0.79 & 0.62 \nl
& 2.5 & D & D & D & D & D & D & D & D \nl
& & 0.034 & 0.31 & 0.034 & 0.32 & 0.035 & 0.32 & 0.034 & 0.33 \nl
& & 0.77 & 0.55 & 0.77 & 0.56 & 0.76 & 0.56 & 0.77 & 0.57 \nl
& 3.5 & D & C & D & D & D & D & D & D\nl
& & 2.5 & 21.6 & 2.9 & 24.0 & 3.1 & 28.1 & 3.2 & 36.2 \nl
& & 0.64 & 0.021 & 0.73 & 0.27 & 0.74 & 0.34 & 0.76 & 0.44 \nl
\multicolumn{10}{c}{}\nl
3 & 1.5 & D & D & D & D & D & D & D & D \nl
& & 0.014 & 0.026 & 0.014 & 0.026 & 0.014 & 0.026 & 0.014 & 0.026 \nl
& & 0.53 & 0.42 & 0.55 & 0.41 & 0.55 & 0.41 & 0.54 & 0.41 \nl
& 2.5 & D & D & D & D & D & D & D & D \nl
& & 0.28 & 2.2 & 0.29 & 2.7 & 0.29 & 2.9 & 0.29 & 3.0 \nl
& & 0.46 & 0.27 & 0.47 & 0.32 & 0.48 & 0.34 & 0.49 & 0.37 \nl
& 3.5 & C & C & C & C & D & C & D & C \nl
& & 21.5 & 32.1 & 44.4 & 79.8 & 42.3 & 129 & 43.5 & 313 \nl
& & 0.078 & 0.17 & 0.035 & 0.14 & 0.085 & 0.13 & 0.28 & 0.11 \nl
\multicolumn{10}{c}{}\nl
7 & 1.5 & D & C & D & C & D & C & D & C \nl
& & 1.0 & 3.1 & 3.0 & 7.7 & 4.2 & 12 & 5.9 & 27 \nl
& & 0.022 & 0.010 & 3.3\sisu{-3} & 0.012
& 8.0\sisu{-3} & 0.013 & 0.023 & 0.011 \nl
& 2.5 & C & C & C & C & C & C & C & C \nl
& & 9.6 & 10.1 & 22.5 & 23.8 & 35.5 & 37.7 & 83.1 & 88.7 \nl
& & 0.26 & 0.32 & 0.26 & 0.32 & 0.26 & 0.32 & 0.25 & 0.31 \nl
& 3.5 & C & C & C & C & C & C & C & C \nl
& & 10.5 & 9.9 & 31.1 & 30.9 & 51.3 & 53.2 & 131.3 & 135 \nl
& & 0.57 & 0.65 & 0.51 & 0.58 & 0.48 & 0.55 & 0.49 & 0.55 \nl
\enddata
\tablenotetext{a}{
The results of Chernoff \& Weinberg (1990, CW) are taken
from their Table 5. Aa represents our resluts for the anisotropic
Fokker-Planck models with the apocenter criterion.}
\tablenotetext{b}{
The first entry describes the fate of the cluster at
the end time of the simulation $t_{\rm end}$:
C (core collapse) or D (disruption).
The second entry is $t_{\rm end}$ in units of $10^9$~yr.
The third entry is the cluster mass at $t_{\rm end}$, $M_{\rm end}$,
in units of the initial mass.}
\end{deluxetable}

\begin{deluxetable}{ccllllll}
\tablecaption{Comparison of CW, Ae, and Aa models for family 1. \label{tab:sur-f1}}
\tablewidth{0pt}
\scriptsize
\tablehead{
\colhead{$W_0$} & \colhead{$\alpha$} & \colhead{Model} &
\colhead{Fate} & \colhead{$t_{\rm end}$} &
\colhead{$M_{\rm end}$} \phantom{xx} &
\colhead{$t(M_{\rm end,CW})$} &
\colhead{$M(t_{\rm end,CW})$} \\
&&&& [Gyr] & [$M_0$] & [Gyr] & [$M_0$]
}
\startdata
1 & 1.5 & CW & D & 0.0092 & 0.78 & 0.0092 & 0.78 \nl
  &     & Ae & D & 0.013 & 0.69 & 0.012 & 0.98 \nl
  &     & Aa & D & 0.017 & 0.61 & 0.014 & 0.98 \nl
  & 2.5 & CW & D & 0.034 & 0.77 & 0.034 & 0.77 \nl
  &     & Ae & D & 0.056 & 0.61 & 0.046 & 0.85 \nl
  &     & Aa & D & 0.31 & 0.55 & 0.12 & 0.89 \nl
  & 3.5 & CW & D & 2.5 & 0.64 & 2.5 & 0.64 \nl
  &     & Ae & D & 3.1 & 0.45 & 2.3 & 0.60 \nl
  &     & Aa & C & 21.6 & 0.021 & 6.3 & 0.82 \nl
3 & 1.5 & CW & D & 0.014 & 0.53 & 0.014 & 0.53 \nl
  &     & Ae & D & 0.019 & 0.49 & 0.019 & 0.76 \nl
  &     & Aa & D & 0.026 & 0.42 & 0.025 & 0.82 \nl
  & 2.5 & CW & D & 0.28  & 0.46 & 0.28 & 0.46 \nl
  &     & Ae & D & 0.45  & 0.38 & 0.40 & 0.58 \nl
  &     & Aa & D & 2.2  & 0.27 & 1.5 & 0.72 \nl
  & 3.5 & CW & D & 21.5 & 0.078 & 21.5 & 0.078 \nl
  &     & Ae & C & 23.6 & 0.097 & -- & 0.14 \nl
  &     & Aa & C & 32.1 & 0.17 & -- & 0.37 \nl
7 & 1.5 & CW & D & 1.0 & 0.022 & 1.0 & 0.022 \nl
  &     & Ae & D & 1.1 & 0.030 & 1.1 & 0.046 \nl
  &     & Aa & C & 3.1 & 0.010 & 2.7 & 0.12 \nl
  & 2.5 & CW & C & 9.6 & 0.26 & 9.6 & 0.26 \nl
  &     & Ae & C & 9.5 & 0.27 & -- & -- \nl
  &     & Aa & C & 10.1 & 0.32 & -- & 0.34 \nl
  & 3.5 & CW & C & 10.5 & 0.57 & 10.5 & 0.57 \nl
  &     & Ae & C & 9.7 & 0.61 & -- & -- \nl
  &     & Aa & C & 9.9 & 0.65 & -- & -- \nl
\enddata
\tablenotetext{a}{Most of the notation is defined in the notes to 
Table \ref{tab:sur}.}
\tablenotetext{b}{
Ae represents anisotropic \FP\ models with the energy criterion.}
\tablenotetext{c}{$M_{\rm end,CW}$ and $t_{\rm end,CW}$ denote 
the end mass and the end time, respectively, of each of the CW models. 
See text for more details.}
\end{deluxetable}

\begin{deluxetable}{cclllll}
\tablecaption{Comparison with \nbody\ models of FH and AH for family 1. \label{tab:fhah}}
\tablewidth{0pt}
\scriptsize
\tablehead{
\colhead{$W_0$} & \colhead{$\alpha$} & \colhead{Model} &
\colhead{$t_{\rm cc}$} & \colhead{$t_{\rm dis}$} &
\colhead{$M_{\rm cc}$} & \colhead{$M_{\rm dis}$} \\
&&& [Gyr] & [Gyr] & [$M_0$] & [$M_0$]
}
\startdata
1 & 1.5 & FH3.7 & -- & 0.076 & -- & 0  \nl
  &     & Aa & -- & 0.017 & -- & 0.61  \nl
  & 2.5 & FH4.0 & -- & 0.29 & -- & 0  \nl
  &     & Aa & -- & 0.31 & -- & 0.55  \nl
  & 3.5 & FH4.1 & -- & 2.2 & -- & 0  \nl
  &     & Aa & 22 & -- & 0.021 & --  \nl
3 & 1.5 & FH3.7 & -- & 0.11 & -- & 0  \nl
  &     & AH3.7 & -- & 0.075 & -- & 0  \nl
  &     & AH9.2 & -- & 0.12 & -- & 0  \nl
  &     & Aa & -- & 0.026 & -- & 0.42  \nl
  & 2.5 & FH4.0 & -- & 0.91  & -- & 0  \nl
  &     & AH4.0 & -- & 3.0  & -- & 0  \nl
  &     & AH10.0 & -- & 3.4  & -- & 0  \nl
  &     & Aa & -- & 2.2  & -- & 0.27  \nl
  & 3.5 & AH4.1 & -- & $>$20 & -- & --  \nl
  &     & AH10.4 & -- & $>$20 & -- & --  \nl
  &     & Aa & 32 & -- & 0.17 & -- \nl
7 & 1.5 & AH3.7 & -- & 1.2 & -- & 0 \nl
  &     & AH9.2 & -- & 3.2 & -- & 0 \nl
  &     & Aa & 3.1 & -- & 0.010 & -- \nl
  & 2.5 & AH4.0 & 13 & $>$20   & 0.21 & -- \nl
  &     & AH10.0 & 11 & $>$20   & 0.257 & -- \nl
  &     & Aa & 10 & -- & 0.32 & -- \nl
  & 3.5 & AH4.1 & 10.4 & $>$20   & 0.606 & -- \nl
  &     & AH10.4 & 10.5 & $>$20   & 0.62 & -- \nl
  &     & Aa & 9.9 & -- & 0.65 & -- \nl
\enddata
\tablenotetext{a}{Part of the notation is defined in the notes to 
Table \ref{tab:sur}.}
\tablenotetext{b}{FH denotes \nbody\ models of Fukushige \& Heggie (1995),
and AH denotes \nbody\ models of Aarseth \& Heggie (1998).
A number associated with each of these model names indicates the
galactocentric
distance $R_{\rm g}$ (in kpc) at which that model is placed.}
\tablenotetext{c}{$t_{\rm cc}$ and $M_{\rm cc}$ denote the collapse time
and mass, respectively. $t_{\rm dis}$ and $M_{\rm dis}$ denote
the disruption time and mass, respectively.}
\end{deluxetable}

\begin{thebibliography}{9999}

\bibitem[]{ah98}
Aarseth, S.\ J., \& Heggie, D.\ C.\ 1998, MNRAS, 297, 794
\bibitem[]{bt87} Binney, J., \& Tremaine, S. 1987, Galactic Dynamics
(Princeton University Press, Princeton)
\bibitem[]{cvgr99} Carraro, G., Vallenari, A., Girardi, L., \& Richichi, A.,
1999, \aap, 343, 825
\bibitem[]{ch93} Chernoff, D. F. 1993, in Structure and Dynamics of
Globular Clusters, ASP Conference Series Vol. 50, 
ed. S. G. Djorgovski \& G. Meylan (San Francisco: ASP), 245
\bibitem[]{cd89} Chernoff, D. F., \& Djorgovski, S. 1989, \apj, 339, 904
\bibitem[]{cw90}
Chernoff, D. F., \& Weinberg, M. D. 1990, ApJ, 351, 121 (CW)
\bibitem[]{c79} Cohn, H.\ 1979, ApJ, 234, 1036
\bibitem[]{c80} Cohn, H.\ 1980, ApJ, 242, 765
\bibitem[]{detal96} Dauphole, B., Geffert, M., Colin, J., Ducourant, C., Odenkirchen, M., \& Tucholke, H. J. 1996, \aap, 313, 119
\bibitem[]{dm99} De Marchi, G., Leibundgut, B., Paresce, F., \& Pulone, L. 1999,
A\&A, 343, L9
\bibitem[]{1995A&A...304..202D} De Marchi, G., \& Paresce, F.\ 1995a, A\&A, 304, 202
\bibitem[]{1995A&A...304..211D} De Marchi, G., \& Paresce, F.\ 1995b, A\&A, 304, 211
\bibitem[]{dk86} Djorgovski, S., \& King, I. R. 1986, \apjl, 305, L61
\bibitem[]{d95} Drukier, G. A., 1995, \apjs, 100, 347
\bibitem[]{d96} Drukier, G. A., 1996, \mnras, 280, 498
\bibitem[]{dcly99} Drukier, G. A., Cohn, H. N., Lugger, P. M., \& Yong, H.
1999, \apj, 518, 233
\bibitem[]{fh95}
Fukushige, T., \& Heggie, D.~C. 1995, MNRAS, 276, 206
\bibitem[]{g88} Giersz, M. 1998, MNRAS, 298, 1239
\bibitem[]{gh94a} Giersz, M., \& Heggie, D.\ C.\ 1994a, MNRAS, 268, 257
\bibitem[]{gh94b} Giersz, M., \& Heggie, D.\ C.\ 1994b, MNRAS, 270, 29
\bibitem[]{gh97} Giersz, M., \& Heggie, D. C. 1997, MNRAS, 286, 709
\bibitem[]{gs94} Giersz, M., \& Spurzem, R.\ 1994,      MNRAS, 269, 241
\bibitem[]{go97} Gnedin, O., \& Ostriker, J. P. 1997, \apj, 474, 223
\bibitem[]{hgst99} Heggie, D.\ C., Giersz, M., Spurzem, R.,
\& Takahashi, K.\ 1999,
in Highlights of Astronomy, Vol. 11, ed. J. Andersen
\bibitem[]{hh96} Heggie, D.\ C., \& Hut, P.\ 1996,
in IAU Symp.~174, Dynamical Evolution of Star Clusters, 
ed. P.~Hut \& J.~Makino (Dordrecht: Kluwer), 303
\bibitem[]{1995MNRAS.272..317H} Heggie, D.\ C., \& Ramamani, N., 1995, MNRAS 272, 317
\bibitem[]{h80} Hills, J. G. 1980, \apj, 225, 986
\bibitem[]{hut94} Hut, P. 1994, in IAU Symp.~165, Compact stars in binaries, 
ed.~ J.\ van Paradijs and E.\ P.\ J.\ van den Heuvel \& E. Kuulkers
(Kluwer), 377
\bibitem[]{hmm95} Hut, P., Makino, J., \& McMillan, S. 1995, ApJ, 443, 93
\bibitem[]{jp94} Janes, K. A., \& Phelps, R. L. 1994, \aj, 108, 1773
\bibitem[]{k66} King, I. 1966, AJ, 71, 64.
\bibitem[]{1990MNRAS.244...76K} Kroupa, P. Tout, C.\ A., \& Gilmore,
G. 1990 MNRAS, 244, 76
\bibitem[]{lg95} Lee, H.\ M., Goodman, J.\ 1995, ApJ, 443, 109
\bibitem[]{lo87} Lee, H.\ M., \& Ostriker, J.\ P.\ 1987, ApJ, 322, 123
\bibitem[]{ma92} Makino, J., \& Aarseth, S.~J. 1992, PASJ, 44, 141
\bibitem[]{1997ApJ...480..432M}
Makino, J., Taiji, M., Ebisuzaki, T., \& Sugimoto, D. 1997, ApJ, 480,
432
\bibitem[]{1991ApJ...369..200M}
Makino, J. 1991, ApJ 369, 200
\bibitem[]{1998MNRAS.293..479M}
Marconi, G., Buonanno, R., Carretta, E., Ferraro, F.\ R.,
        Fusi Pecci, F., Montegriffo, P., De Marchi, G.,
        Paresce, F., \& Laget, M. 1998, MNRAS, 293, 479
\bibitem[]{1986ApJ...307..126M}McMillan, S. L. W. 1986a, ApJ 307, 126
\bibitem[]{1986ApJ...306..552M}McMillan, S. L. W. 1986b, ApJ 306, 552
\bibitem[]{mh96}McMillan, S.\ L.\ W., Hut, P. 1996 ApJ 467, 348
\bibitem[]{mh97} Meylan, G., \& Heggie, D.\ C. 1997, \aapr\, 8,1
\bibitem[]{mw97c} Murali, C., \& Weinberg, M. D. 1997, \mnras, 291, 717
\bibitem[]{p97} Phelps, R. L. 1997, \apj, 483, 826
\bibitem[]{pzv96}
Portegies~Zwart, S.~F., \& Verbunt, F. 1996, A\&A, 309 179
\bibitem[]{pzy98}
Portegies Zwart, S.\ F., \& Yungelson, L.\ 1998, A\&A, 332, 173
\bibitem[]{1998A&A...337..363P}
Portegies Zwart, S.\ F., Hut, P., Makino, J., \& McMillan, S.\ L.\ W.\ 1998, 
A\&A, 337, 363
\bibitem[]{pzvdh99} Portegies Zwart S.F., van den Heuvel E. P. J.,
1999, New Astron., 4, 355 
\bibitem[]{pm93} Pryor, C., \& Meylan, G. 1993, 
in Structure and Dynamics of
Globular Clusters, ASP Conference Series Vol. 50,
ed. S. G. Djorgovski \& G. Meylan (San Francisco: ASP), p357
\bibitem[]{q96} Quinlan, G. D., 1996, New Astron., 1, 255
\bibitem[]{rmh97} Ross, D.\ J., Mennim, A., \& Heggie, D.\ C.\ 1997, MNRAS, 284, 811
\bibitem[]{sca86} Scalo, J.\ M.\ 1986, Fund. of Cosm. Phys., 11, 1
\bibitem[]{sp87} Spitzer, L. Jr. 1987, Dynamical Evolution of Globular
Clusters (Princeton University Press, Princeton)
\bibitem[]{st95} Spurzem, R., \& Takahashi, K.\ 1995, MNRAS, 272, 772
\bibitem[]{t95}
Takahashi, K.\ 1995, PASJ, 47, 561
\bibitem[]{t96}
Takahashi, K.\ 1996, PASJ, 48, 691
\bibitem[]{t97}
Takahashi, K.\ 1997, PASJ, 49, 547
\bibitem[]{tl99}
Takahashi, K., \& Lee, H. M., 1999, submitted to MNRAS (astro-ph/9909006)
\bibitem[]{tli97}
Takahashi, K., Lee, H.~M., \& Inagaki, S.\ 1997, MNRAS, 292, 331
\bibitem[]{ktpz98}
Takahashi, K., \& Portegies Zwart, S.\ F.\ 1998, ApJ, 503, L49 (TPZ)
\bibitem[]{tdk93} Trager, S. C., Djorgovski, S., \& King, I. R. 1993, 
in Structure and Dynamics of
Globular Clusters, ASP Conference Series Vol. 50,
ed. S. G. Djorgovski \& G. Meylan (San Francisco: ASP), 347
\bibitem[]{vh97} Vesperini, E., \& Heggie, D. C. 1997, MNRAS, 289, 898


\end{thebibliography}
\end{document}